\documentclass[preprint,12pt,apj]{emulateapj}
\usepackage{graphicx}
\usepackage{times}
\usepackage{color}
\usepackage{natbib}
\usepackage{cancel}
\usepackage{float}
\usepackage{epsfig,epsf}
\usepackage{bm,amssymb,amsmath}
  \newcommand{\Mm}{\,{\rm Mm}}
  \newcommand{\km}{\,{\rm km}}

  \newcommand{\K}{\,{\rm K}}
  
  \newcommand{\G}{\,{\rm G}}
  
  \newcommand{\mT}{\,{\rm mT}}

  \newcommand{\mkg}{\,\mu{\rm g}}

  \newcommand{\mcube}{\,{\rm m^{-3}}}
  \usepackage{bm,amssymb,amsmath}
  \newcommand{\vect}[1]{{{\mbox{\boldmath $#1$}}}}		
  \newcommand{\GO}{{G}}
  \newcommand{\bc}{{b_{00}}}
  \newcommand{\BO}{{B_{0z}}}
  \newcommand{\zh}{{\ell}}
  \newcommand{\fO}{{f_{0}}}
  \newcommand{\bF}{{b_{01}}}
  \newcommand{\za}{{z_{1}}}
  \newcommand{\bb}{{b_{02}}}
  \newcommand{\zb}{{z_{2}}}
  \newcommand{\vv}{{_{\rm h}}}  				
  \newcommand{\hh}{{_{\rm m}}}  				
  \newcommand{\dd}{\,{\rm d}}       				
  \newcommand\leftidx[3]{%
  {\vphantom{#2}}\,#1\hspace{-0.05cm}#2#3%
}
  \newcommand{\GFME}{{Paper~I}} 
  \definecolor{burntorange}{RGB}{255,97,0}
  \newenvironment{apjrev}{\color{black}}{}
  \shorttitle{Flux tube magnetohydrostatic equilibrium}
  \shortauthors{Gent, Fedun \& Erd\'{e}lyi}
\begin{document}
  \title{\MakeUppercase{Magnetohydrostatic equilibrium. II. Three-dimensional multiple open
         magnetic flux tubes in the stratified solar atmosphere}}

  \author{F.~A.~Gent$^{1}$, V.~Fedun$^{2}$ and R.~Erd\'{e}lyi$^1$}
  \affil{$^1$SP$^2$RC, School of Mathematics and Statistics, University of 
            Sheffield, S3 7RH, UK: f.gent@shef.ac.uk} 
  \affil{$^2$Space Systems Laboratory, Department of Automatic Control and 
           Systems Engineering, University of Sheffield, S1 3JD, UK}
    
  \begin{abstract}
    
    A system of multiple open magnetic flux tubes spanning the solar 
    photosphere and lower corona is modeled analytically, within a 
    realistic stratified atmosphere subject to solar gravity. 
    This extends results for a single magnetic flux tube in 
    magnetohydrostatic equilibrium, described in Gent et al. (MNRAS,
    {\bf{435}}, 689, 2013). 
    Self-similar magnetic flux tubes are combined to form magnetic
    structures, which are consistent with high-resolution observations.
    The observational evidence supports the existence of strands of
    open flux tubes and loops persisting in a relatively steady state.
    Self-similar magnetic flux tubes, for which an analytic solution
    to the plasma density and pressure distribution is possible, are combined. 
    We calculate the appropriate balancing forces, applying to the equations
    of momentum and energy conservation to preserve equilibrium.

    Multiplex flux tube configurations are observed to remain relatively stable 
    for up to a day  or more, and it is our aim to apply our model as the
    background condition for numerical studies of energy transport mechanisms 
    from the solar surface to the corona. 
    We apply magnetic field strength,
    plasma density, pressure and temperature distributions consistent with 
    observational and theoretical estimates for the lower solar atmosphere.
    Although each flux tube is identical in construction apart from the 
    location of the radial axis, combinations can be applied to generate
    a non-axisymmetric magnetic field with multiple non-uniform flux tubes.
    This is a considerable step forward in modeling the realistic
    magnetized three-dimensional equilibria of the solar atmosphere.
    
  \end{abstract}


  \section{Introduction}\label{Intro}
  At a radius $R_\odot\simeq696\Mm$ from the Sun's core its luminous 
  surface, the photosphere, has a {\apjrev{typical}} temperature of about 
  $6500\K$. 
  {\apjrev{Based on estimates from semi-empirical 1D models a}}t 
  $h\simeq0.35$--$0.65\Mm$ above this surface, the temperature falls to a
  minimum $T\simeq4200\K$. 
  The temperature then rises with height and experiences rapid jumps
  to $10^5\K$ just above $h\simeq2\Mm$ and to $10^6\K$ beyond 
  $h\simeq2.5\Mm$ 
  \citep[][and references therein]{VAL81,Priest87,Asch05,Ebook08}.
  The mechanism {\apjrev{of}} heating the solar corona is not well understood. 
  The solar atmosphere is highly active. 
  Jets, flares, prominences{\apjrev{, spicules and flux emergence, among 
  others,}} carry mass and energy from the surface into the atmosphere. 
  Although frequent and powerful, these solar accumulated events do not yet 
  appear to be of sufficient energy to explain the consistently high 
  temperatures for the 
  corona {\apjrev{\citep{Zirker93,Asch05,Klimchuk06}}}.  
  An alternative view may be that solar magnetic field lines, in the form of
  magnetic \emph{flux tubes}, act as guides for magnetohydrodynamic (MHD) waves
  that may carry the missing energy to heat the corona to observed 
  temperatures \citep[]{JMMKAB07,MVJKRME12,WSSRLFE12}

  Coronal loops, comprising strongly magnetized {flux tubes}, permeate the
  atmosphere. 
  Given the very low thermal pressure in the solar corona, the
  magnetic pressure can become dynamically dominant.
  From the photosphere to the lower corona, there is a drop of six orders
  of magnitude in the plasma pressure and nine orders of magnitude in the 
  plasma density \citep{VAL81}. 
  Just above $2\Mm$ from the photosphere there is a transition zone, called the
  Transition Region (TR), where
  there is a jump in plasma density and temperature of two orders of
  magnitude, evident in the black lines of Figure\,\ref{fig:hydroz}
  {{\citep[see also Figure\,1 of][hereafter referred
   to as \GFME]{GFME13a}.}}

  Typical footpoint strengths of $100\mT~ (1000\G)$ are observed for magnetic
  flux tubes emerging from the photosphere 
  \citep[][and references therein, the latter Chapter 8.7, Chapter 5, 
  respectively]{Zwaan78,Priest87,Asch05,
  Ebook08}.
  An isolated magnetic flux tube must therefore expand exponentially in radius
  as it approaches 
  the TR to balance the plasma pressure.
  Environments with such large dynamical scales are highly challenging to
  model \citep{Def07}.

  Although on some solar timescales they may be regarded as transient 
  features, magnetic flux loops persist in relative pressure
  equilibrium with the ambient atmosphere for many minutes, days or longer
  {\apjrev{\citep{MTKWSS77,LW77,MSRM83}}}.
  Let us consider the magnetic field as a wave guide for carrying 
  energy from the lower solar atmosphere and releasing it as heat high in the
  corona.
  We can take advantage of the steady background state of the magnetic field 
  and plasma to investigate such transport mechanisms with a series of 
  numerical  
  simulations {{\citep{SFE08,FES09,SZFET09,FSE11,VFHE12,KC12,MFE14}}}.

  Magnetic flux tubes appear to exhibit overdense cores in the corona,
  in apparent contradiction with hydrostatic equilibrium \citep{ASA01,WWM03}.
  Modeling a single flux tube in pressure equilibrium for the corona 
  dictates that the internal magnetic pressure arising from a predominantly
  parallel field 
  will reduce the plasma  pressure and, consequently, also plasma density or 
  temperature.
  Combining multiple flux tubes may induce magnetic tension forces 
  restoring and potentially even enhancing plasma density within the flux
  tubes.
  Figure\,\ref{fig:hydroz}(a) displays the axial profile for the plasma
  pressure, temperature and density with the same 
  parameters for the magnetic flux tube as applied in \GFME, but
  with some revisions as outlined in Section~\ref{sect:single}.
  Significantly, in the corona, the plasma density inside the flux 
  tube is lower than the ambient plasma. 
  Note that a more gradual expansion of the flux tube is applied (panel (b), 
  below $1\Mm$, blue, dashed line). 
  The resulting plasma
  density, and, to some extent, the pressure, is enhanced in the chromosphere
  and TR, where there are strong tension forces applying, 
  but not in the corona where the field lines are predominantly parallel.
  \begin{center}
  \includegraphics[width=0.9\linewidth]{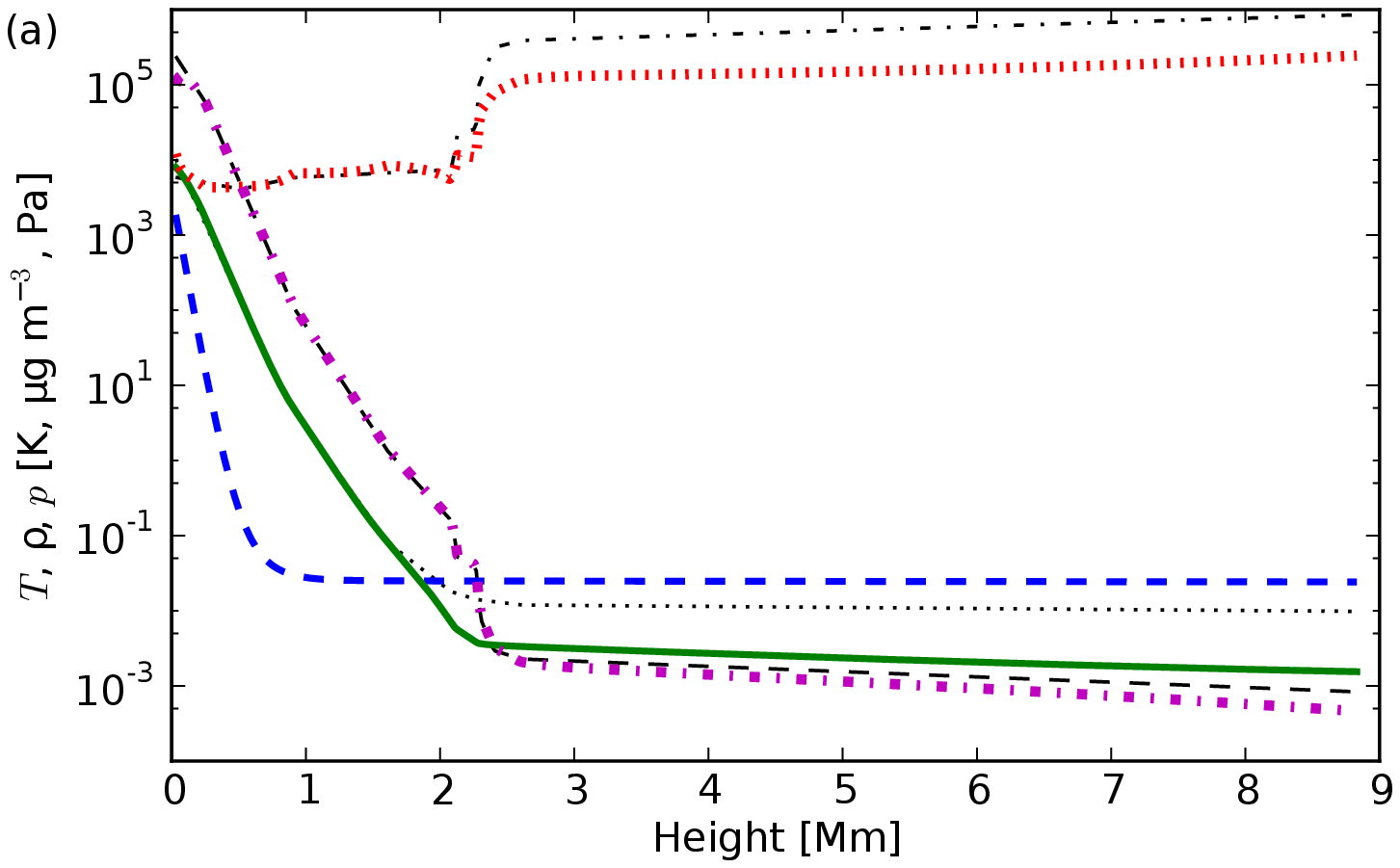}
  \includegraphics[width=0.9\linewidth]{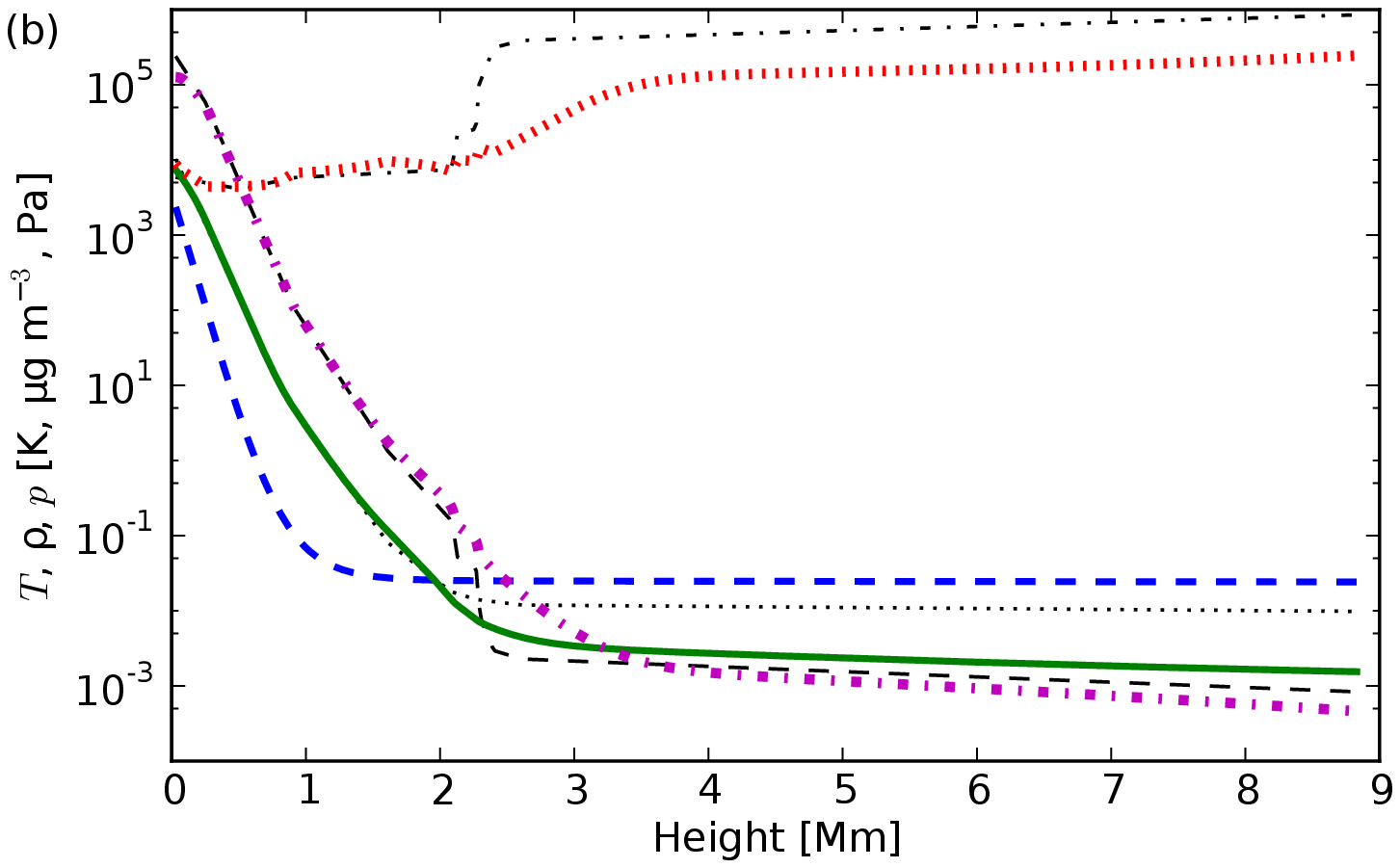}
  \figcaption{
  1D-slices along the axis of a single magnetic flux tube for 
  thermal $p$\,[Pa] (solid, green) and 
  magnetic $B^2/(2\mu_0)$\,[Pa] (dashed, blue) pressure, 
  temperature $T\,[\K]$ (dotted, red) 
  and 
  plasma density $\rho\,[\mkg\mcube]$ (dash-dotted, purple).
  Panel {(a)} has parameters matching \GFME, while {(b)}
  applies a reduced expansion rate for the flux tube in the chromosphere.
  Interpolated 1D fits to the vertical atmospheric profile are shown in black 
  lines \citep[][former up to 2.3\,Mm; 
  latter above 2.4\,Mm]{VAL81,MTW75}. 
  Differences between model and reference profiles vanish
  away from the flux tube axis.
  \label{fig:hydroz}}
  \end{center}
 
  Numerical models with a single flux tube may miss some of the  
  interesting non-linear effects arising from interactions between 
  neighboring flux tubes.
  \citet{KCF08,KC12} have constructed a two-dimensional (2D) magnetic field with
  multiple flux
  tubes for a domain which does not include the transition region and where 
  each flux tube is identical to its neighbor. 
  \citet{HvBKS05,HvB08} have constructed a 2D magnetic field which does 
  extend into the low corona.
  In this paper, in a domain from the photosphere to beyond the TR, we 
  construct a three-dimensional (3D) magnetic field with inhomogeneous multiple flux tubes.
  This is a considerable step forward in realistic modeling
  of 3D magnetic networks embedded in the highly stratified solar atmosphere.
 
  In a series of papers, 
  \citet[e.g.,][]{BCLow85,BCLow88} 
  describes a
  method for deriving analytically the equilibrium plasma 
  pressure and density distribution for a set of magnetic field configurations.
  This result was applied to the global solar magnetic field structure and
  coronal mass ejections \citep[e.g.,][]{TL89,GL98}.

  Here, we describe an alternative, empirical
  method for constructing an equilibrium magnetic field 
  comprising multiple non-uniform flux tubes 
  within a realistic stratified solar atmosphere.
  Our aim is to analytically describe a structure for the magnetic
  field, matching observational models \citep[e.g.,\,][]{LDK08,KHM08,VGH11,JK13}.
  We derive analytic expressions for adjustments to the plasma pressure and
  density, due to the magnetic field, to restore pressure 
  balance, also constrained by observational parameters.
  We empirically identify the minimal balancing forces applying 
  to the MHD 
  equations of momentum and energy conservation that preserve this equilibrium.

  The paper is organized as follows. 
  Section\,\ref{subsect:strat} clarifies what changes have been introduced
  in comparison to the single magnetic flux tube model detailed in 
  \citet{GFME13a}. 
  Section\,\ref{subsect:MHD} describes how the MHD equations governing the
  perturbed system must be framed to account for the steady background 
  utili{\apjrev{z}}ing multiple flux tubes of the form defined in 
  Section\,\ref{subsect:mag}.  
  Section\,\ref{subsect:initHD} outlines
  how the atmosphere is adjusted to balance the magnetic pressure and tension
  forces and identifies
  the balancing forces which must also be applied.
  In Section\,\ref{sect:change}, the changes to the MHD equations
  are identified.
  Section\,\ref{sect:multi} has examples of heterogeneous
  multiple flux tube fields that are possible with this method.
  A summary of the results are presented in Section\,\ref{sect:conc}
  with some points for discussion.
  In the Appendix, we show some of the analytic 
  calculations in more detail.
  For pairs of interacting magnetic flux tubes
  Appendix\,\ref{sect:curl} demonstrates why a balancing force must be
  present in addition to any changes to the
  plasma pressure and density, and identifies these forces.
  Profiles for the 
  plasma pressure and density are derived in Appendix\,\ref{sect:pbal}.
  
  \section{Multiple open magnetic flux tubes}\label{sect:single}
  \subsection{Development Beyond the Single Flux Tube Model}\label{subsect:strat}
%
  Following \citet{GFME13a} we apply a background atmosphere 
  derived by a combination of modeling profiles from 
  \citet[][Table~12, VALIIIC]{VAL81}
  and \citet[][Table~3]{MTW75} for the chromosphere and lower solar 
  corona, respectively,
  assuming background equilibrium parameters for the quiet Sun.
  The profiles interpolated as a function of height above the
  photosphere are included in Figure\,\ref{fig:hydroz}
  as black dotted lines (pressure), dashed (density), and dash-dotted 
  (temperature).
  Up to the
  TR, at around $2.2\Mm$, 
  the steep pressure and corresponding density gradients 
  dictate that magnetic flux tubes expand
  rapidly in radius and reduce in flux density.
  In the solar corona, the flux tube radius is almost steady with height.
  Our models, both for the single tube and the multiple configurations,
  capture the reference data profiles very effectively.
  However, the models do not depend on the choice of atmosphere and the
  derivation described could be applied to many alternative atmospheric
  models. 
 
  In constructing the magnetic field, 
  we include a
  constituent to represent an ambient magnetic field, ubiquitous within
  the solar atmosphere. 
  In \GFME\, the ambient field was a function of $r$ and $z$.
  Here we apply a constant vertical ambient field, which 
  still satisfies the divergence free condition and, as constructed, 
  retains thermal pressure $p>0$ as $z\rightarrow\infty$.
  The resulting derivatives therefore somewhat simplify, compared to \GFME.
  
  For an axially symmetric flux tube, it is convenient to work in polar
  coordinates. 
  This is applied in \GFME\, and for the individual flux tube 
  element of Appendix\,\ref{sect:pbal} in this paper.
  For a pair of flux tubes that differ by axial coordinates, however, axial
  symmetry is broken. 
  In this paper, it is therefore, more convenient to compute the
  flux tube interactions in Cartesian coordinates.

  In Equation\,\eqref{eq:momb} for pressure balance the magnetic 
  tension force is of opposite sign to the gradients of pressure. 
  In \GFME\, the tension force has the wrong sign.
  The derivations remain valid.
  With the sign corrected, the flux tube axial
  plasma density is no longer enhanced in the corona, although it is in the 
  chromosphere, where the tension forces are strongest.
  This suggests that the enhanced density observed in steady flux tubes may be
  due to the magnetic tension forces governed by the curvature of the 
  magnetic field lines.

  With the pressure and tension forces correctly in opposition, there
  is much more latitude in the model parameters. 
  The thickness of the flux tube can 
  extend to over $1\Mm$, with an upper bound on the footpoint field strength
  near $150\mT$ without negative pressure or density resulting.
  Solar magnetic flux tubes are unlikely to exhibit such homogeneity.
  Combining many small uniform flux tubes in non-uniform distribution, we
  may construct large heterogeneous flux tubes.
  
\subsection{The MHD Equations} \label{subsect:MHD}
  A motivation for the current work is to facilitate an MHD solution in an
  environment that includes a plasma density gradient with nine orders of 
  magnitude over a relatively short vertical span. 
  By confining this and other large gradients within a static background,
  the MHD equations can be re-framed with respect to the significantly more
  modest differentials of the perturbations.

  {\apjrev{This article employs several subscripts and superscripts. 
  Subscripts $i,j\in\{1,2,3\}$ denote general
  vector or tensor coordinate components only.
  The convention of a sum over all three components applies for a repeated
  index in a single expression.
  In Cartesian coordinates $x_i = (x_1,x_2,x_3) = (x,y,z)$, and where 
  $x,\,y$ or $z$ appear as subscripts they refer only to the respective
  component $x_1,\,x_2$ or $x_3$. 
}}

  The governing equations of full ideal compressible MHD in their conservative
  form are:
  \begin{eqnarray}
    \label{eq:mass}
    \frac{\partial \rho}{\partial t} + 
    \frac{\partial}{\partial x_i}(\rho u_i)& =& 0,\\
    \label{eq:mom}
    \frac{\partial (\rho u_i)}{\partial t}
    +\frac{\partial}{\partial x_j}
    \left(
    \rho u_iu_j
    -\frac{B_iB_j}{\mu_0}
    \right)
    +\frac{\partial p_T}{\partial x_i}
    & =& \rho g_i,\\
    \label{eq:energy}
    \frac{\partial e}{\partial t}
    +\frac{\partial}{\partial x_i}
    \left(e u_i 
    +p_Tu_i \right)
    -\frac{\partial}{\partial x_j}
    \left( 
    \frac{B_iB_ju_i}{\mu_0}
    \right)
    & =& \rho g_i u_i,\\
    \label{eq:ind}
    \frac{\partial B_i}{\partial t} 
    +\frac{\partial }{\partial x_j}
    \left(
    u_iB_j-u_jB_i
    \right)
    &=& 0_i,
  \end{eqnarray}
  \begin{eqnarray}
    \label{eq:pkin}
   p_T = p + \frac{B_jB_j}{2\mu_0},\,\, 
    p = (\gamma-1)
    \left(
     e-\frac{\rho u_ju_j}{2}-\frac{B_jB_j}{2\mu_0}
    \right),\,\,
  \end{eqnarray}
  where $\rho$ is plasma density, and $\nabla,\,\vect{u},\,\vect{B}$ and 
  $\vect{g}$ are the gradient, velocity, magnetic field, and gravitational 
  acceleration vectors, respectively. 
  $e$ is the total energy density, $p$ is the
  thermal pressure, $p_T$ is the total pressure (magnetic + thermal), and
  $\gamma$ is the adiabatic index of the plasma and $\mu_0$ vacuum 
  magnetic permeability.

  Following the approach of \citet{SFE08}, we derive the system of
  equations governing the perturbed MHD variables.
  The variables $\rho$, $e$, and $\vect{B}$ are split into their background 
  and perturbed components
  \begin{equation}
    \label{eq:tilde}
    \rho = \rho_b + {\tilde{\rho}},
\quad    e = e_b + {\tilde{e}},
\quad    \vect{B} = \vect{B}_b + \tilde{\vect{B}},
  \end{equation}   
  where the tilde denotes the perturbed portion and it is assumed $\rho_b,\,e_b$
  and $\vect{B}_b$ do not vary with time.
  {\apjrev{The subscript $b$ denotes background, and in combination with $i$ 
  or $j$, later in the paper, 
  indicates the background vector component.
  }}With magnetohydrostatic equilibrium for the background state, such that
  $\vect{u}_b=\vect{0}$, in the presence of an external gravity field 
  $\vect{g}$, from Equation\,\eqref{eq:mom} we obtain 
  \begin{equation}
    \label{eq:momb}
    \frac{\partial}{\partial x_i}
    \left(
    p_b + \frac{B_{bj}B_{bj}}{2\mu_0}
    \right)
    -\frac{\partial}{\partial x_j}
    \left(
    \frac{B_{bi}B_{bj}}{\mu_0}
    \right)
     = \rho_b g_i.
  \end{equation}
  We then obtain the expression matching the right-hand side of 
  Equation\,\eqref{eq:energy} by scalarly multiplying Equation\,\eqref{eq:momb} by 
  $\vect{u}$ to yield
  \begin{equation}
    \label{eq:energyb}
    u_i\frac{\partial}{\partial x_i}
    \left(
    p_b + \frac{B_{bj}B_{bj}}{2\mu_0}
    \right)
    -u_i\frac{\partial}{\partial x_j}
    \left(
    \frac{B_{bi}B_{bj}}{\mu_0}
    \right)
     = \rho_b g_iu_i.
  \end{equation}
  Subtracting Equations\,\eqref{eq:momb} and \eqref{eq:energyb} from 
  Equations\,\eqref{eq:mom} and \eqref{eq:energy},
  we derive the governing equations
  for the perturbations as
  \begin{eqnarray}
    \label{eq:massp}
    \frac{\partial \tilde\rho}{\partial t} + 
    \frac{\partial}{\partial x_i}[(\rho_b+\tilde\rho) u_i]& =& 0,\\
    \label{eq:momp}
    \frac{\partial [(\rho_b+\tilde\rho) u_i]}{\partial t}
    +\frac{\partial
    }{\partial x_j}
    \left[
    (\rho_b+\tilde\rho) u_iu_j
    \right]
    +\frac{\partial \tilde p_T
    }{\partial x_i}
    &&\nonumber\\
    -\frac{\partial}{\partial x_j}
    \left[
    \frac{\tilde{B_i}B_{bj}+B_{bi}\tilde{B_j}+\tilde{B_i}\tilde{B_j}}{\mu_0}
    \right]
    & =& \tilde\rho g_i,\\
    \label{eq:energyp}
    \frac{\partial \tilde{e}}{\partial t}
    +\frac{\partial}{\partial x_i}
    \left[
    (e_b+\tilde{e}) u_i 
    +\tilde p_T u_i 
    \right]
    &&\nonumber\\
    -\frac{\partial}{\partial x_j}
    \left[
    \frac{\tilde{B_i}B_{bj}+B_{bi}\tilde{B_j}+
     \tilde{B_i}\tilde{B_j}}{\mu_0}u_i
    \right]&&\nonumber\\
    +p_{bT}\frac{\partial u_j}{\partial x_j} 
    - \frac{B_{bj}B_{bi}}{\mu_0} \frac{\partial u_i}{\partial x_j}
     & =&
    \tilde\rho g_i u_i,\\
    \label{eq:indp}
    \frac{\partial \tilde{B_i}}{\partial t} 
    +\frac{\partial }{\partial x_j}
    \left[
    u_i(B_{bj}+\tilde{B_j})-u_j(B_{bi}+\tilde{B_i})
    \right]
    &=& 0_i,
  \end{eqnarray}
  \begin{equation}
    \label{eq:ptotp}
    \tilde p_T = (\gamma-1)
    \left[
     \tilde{e}-\frac{(\rho_b+\tilde\rho) u_ju_j}{2}
    \right]
    -(\gamma-2)
    \left[
     \frac{\tilde{B_j}B_{bj}}{\mu_0}
     +\frac{\tilde B_j\tilde B_j}{2\mu_0}
    \right],
  \end{equation}
  \begin{equation}
    \label{eq:ptotb}
     p_{bT} = (\gamma-1)
     {e}_b
    -(\gamma-2)
     \frac{B_{bj}B_{bj}}{2\mu_0},
  \end{equation}
  \begin{equation}
    \label{eq:pkinp}
    \tilde{p} = (\gamma-1)
    \left[
     \tilde{e}-\frac{(\rho_b+\tilde\rho) u_ju_j}{2}
     -\frac{\tilde{B_j}B_{bj}}{\mu_0}
     -\frac{\tilde B_j\tilde B_j}{2\mu_0}
    \right],
  \end{equation}
  \begin{equation}
    \label{eq:pkinb}
    {p}_b = (\gamma-1)
    \left[
     {e}_b
     -\frac{B_{bj}B_{bj}}{2\mu_0}
    \right].
  \end{equation}

  \subsection{Magnetic Field Construction }\label{subsect:mag}

  Our approach is to prescribe the magnetic field to model a flux tube or
  a set of flux tubes with structure approximating the observed 
  magnetic field in the lower solar atmosphere.
  We place this field in a hydrostatic stratified atmosphere derived from the
  observed vertical profiles of the reference data.
  We then adjust the plasma pressure and density distribution from the 
  hydrostatic background as required to achieve magnetohydrostatic 
  equilibrium.
   
  One approach to constructing the magnetic field is to apply a potential field
  to the prescribed atmosphere and allow the system to relax 
  numerically \citep[e.g.,][]{SS90,KCF08}. 
  Simulations of non-potential perturbations may then be applied 
  to this equilibrium.
  For models utilizing very large data arrays,
  there may be considerable numerical overheads before the simulations
  can proceed, though it may be possible to circumvent this problem
  by using damping methods.
  It is conceivable that we may wish to investigate how small changes to the
  configuration effect energy transport mechanisms.
  With our analytic approach, these changes may be implemented almost 
  instantaneously, and we can identify in advance exactly what changes are 
  applied to the configuration.
  Using the potential method, the preliminary numerical relaxation must be 
  completed and then the change to the configuration investigated.
  It is possible the potential method may be unsuitable for deriving
  background equilibrium multiple flux tube configurations. 
  In this paper, we find no equilibrium exists for neighboring
  pairs of self-similar flux tubes in the absence of balancing forces.
  We are able to identify and calculate these forces. 

  We revisit the self-similarity method
  developed by \citet{ST58} and applied variously for 2D
  \citep[e.g.,][]{Deinzer65,Low80,SR05,GJ07,FSE11,SFKEM11}
  and to 3D solar magnetic configurations \citep{FVJE11,GFME13a,MFE14}.
  This represents one footpoint of a coronal loop or braid of loops.
  The other footpoint is presumed to be at a distance
  beyond the horizontal extent of our numerical domain.
  The arch of the loop occurs much higher in the corona than the vertical 
  extent of our model (i.e., the loop has large aspect ratio),
  such that the flux tube may be regarded as
  vertically aligned. 

  The 3D magnetic field describing 
  the configuration for a single open magnetic flux tube 
  is denoted $\leftidx{^m}{\vect{B}_b}$, {{where $\leftidx{^m}{}$ 
  indicates the $m^{\rm th}$ flux tube in a magnetic field comprising more
  than one flux tube}}.
  To distinguish the index label for each magnetic flux tube 
  configuration from other indices in this article, we use only labels 
  $m,n\in {\mathbb{N}}$ and these appear as prefix superscripts.
  Summation convention does not apply to repetition of these indices.
  The model domain may reasonably be approximated in 
  cylindrical polar coordinates, with radius measured from the axis of the 
  flux tube, or in Cartesian coordinates, with $x,y$ the local analogue of 
  the longitudinal and latitudinal surface coordinates.
  The vertical coordinate $z$ is aligned along the solar radius, with $z=0$ 
  at the base of
  the solar photosphere at $R_\odot\simeq696\Mm$.
  We require an axially symmetric flux tube with its axis located 
  at $(\leftidx{^m}x,\leftidx{^m}y) $,
  expanding in radius with height $z$ as the flux density reduces to balance
  the ambient plasma pressure. 

  In Cartesian coordinates the components of $\leftidx{^m}{\vect{B}_b}$ are 
  described by the self-similar relations
  \begin{align}\label{eq:Bdef} 
   \leftidx{^m}B_{bx}\quad&=
    \hspace{-0.0cm}-\leftidx{^m}S
    \frac{\partial \leftidx{^m}f }{\partial x}
    \frac{\partial \leftidx{^m}f }{\partial z}\leftidx{^m}\GO ,\hspace{2.95cm}
    \\
    \nonumber
    \leftidx{^m}B_{by}\quad&=
    \hspace{-0.0cm}-\leftidx{^m}S
    \frac{\partial \leftidx{^m}f }{\partial y}
    \frac{\partial \leftidx{^m}f }{\partial z}\leftidx{^m}\GO ,\hspace{2.95cm}
    \\
    \nonumber
    \leftidx{^m}B_{bz}\quad&=\hspace{0.3cm}
    \leftidx{^m}S
    \left[
      \left(
        \frac{\partial \leftidx{^m}f }{\partial x}
      \right)^2
      +
      \left(
        \frac{\partial \leftidx{^m}f }{\partial y}
      \right)^2
    \right]
    \leftidx{^m}\GO  + \bc.
  \end{align} 
  The sign $\pm 1$ is indicated by $\leftidx{^m}S$ to determine the orientation
  of the magnetic field along the $m^{\rm th}$ flux tube. 
  Real {{$\bc$ is a constant, chosen to yield a weak ambient vertical 
  field in which any flux tubes are situated.}}
  \begin{align}
    \label{eq:fi}
    \leftidx{^m}f \, & =& \leftidx{^m}r  \BO,
    \\
    \label{eq:GOi}
    \leftidx{^m}\GO \, &=& 
    \frac{2\zh}{\sqrt{\pi}\fO}\exp\left(-\frac{\leftidx{^m}f ^2}{\fO^2}\right),
    \\
    \label{eq:ri}
    \leftidx{^m}r \, &=& \sqrt{(x-\leftidx{^m}x )^2+(y-\leftidx{^m}y )^2},
  \end{align}
  where $\leftidx{^m}r $ is the radial distance from the axis at 
  $(\leftidx{^m}x,\leftidx{^m}y) $ and with 
  $\leftidx{^m}\GO $ determining the radial width of the flux tube by 
  a Gaussian centered at $(\leftidx{^m}x,\leftidx{^m}y) $. 
  In the normalization coefficient, an appropriate length scale $\zh$ is
  included with the scaling factor $\fO$, which are uniform for all flux tubes.
  The reduction in the vertical field strength along the flux tube axis
  is specified by 
  \begin{equation}\label{eq:BO}
    \BO = 
    \bF\exp\left(-\frac{z}{\za}\right)
    +
    \bb\exp\left(-\frac{z}{\zb}\right),
  \end{equation}
  with $\bF$ and $\bb$ assigning the typical axial field strength 
  from the
  photosphere and from the lower corona, respectively.
  $\za$ and $\zb$ are scaling lengths.
  In principle, these constants could differ between each flux tube, yielding
  stronger non-uniformity in the total field.
  Keeping $\BO$ uniform sufficiently simplifies the equations for an
  analytic solution. 
  The final result retains significant asymmetry.

  Applying these to Equation\,\eqref{eq:Bdef} we obtain the explicit form for the
  single magnetic flux tube as
  \begin{align}\label{eq:Bxyz}
    \leftidx{^m}B_{bx}
    &=
    -\leftidx{^m}S(x-\leftidx{^m}x )
    {\BO\leftidx{^m}\GO \, }
    \frac{\partial \BO}{\partial z}
    ,\nonumber
    \\
    \leftidx{^m}B_{by}
    &=
    -\leftidx{^m}S(y-\leftidx{^m}y )
    {\BO\leftidx{^m}\GO  \,}
    \frac{\partial \BO}{\partial z}
    ,
   \nonumber \\
    \leftidx{^m}B_{bz}
    &=
    \phantom{-}\leftidx{^m}S{\BO^2\leftidx{^m}\GO  \,}+\bc.
  \end{align}

  Observations \citep[Chapter 3.5 in][]{M93book,TS03} indicate the atmosphere 
  outside the flux tubes 
  includes a non-zero magnetic field in many parts of the chromosphere and,
  due to the local turbulence, is likely to 
  be composed of very small-scale structures. 
  However, at the scales of interest the structure of this weak field is not
  likely to be dynamically significant.
  For simplicity in satisfying the divergence free
  condition, a vertical magnetic field seems reasonable.
  
\subsection{Plasma Pressure and Density} \label{subsect:initHD}

  \begin{figure*}
  \centering
  \includegraphics[width=0.85\paperwidth]{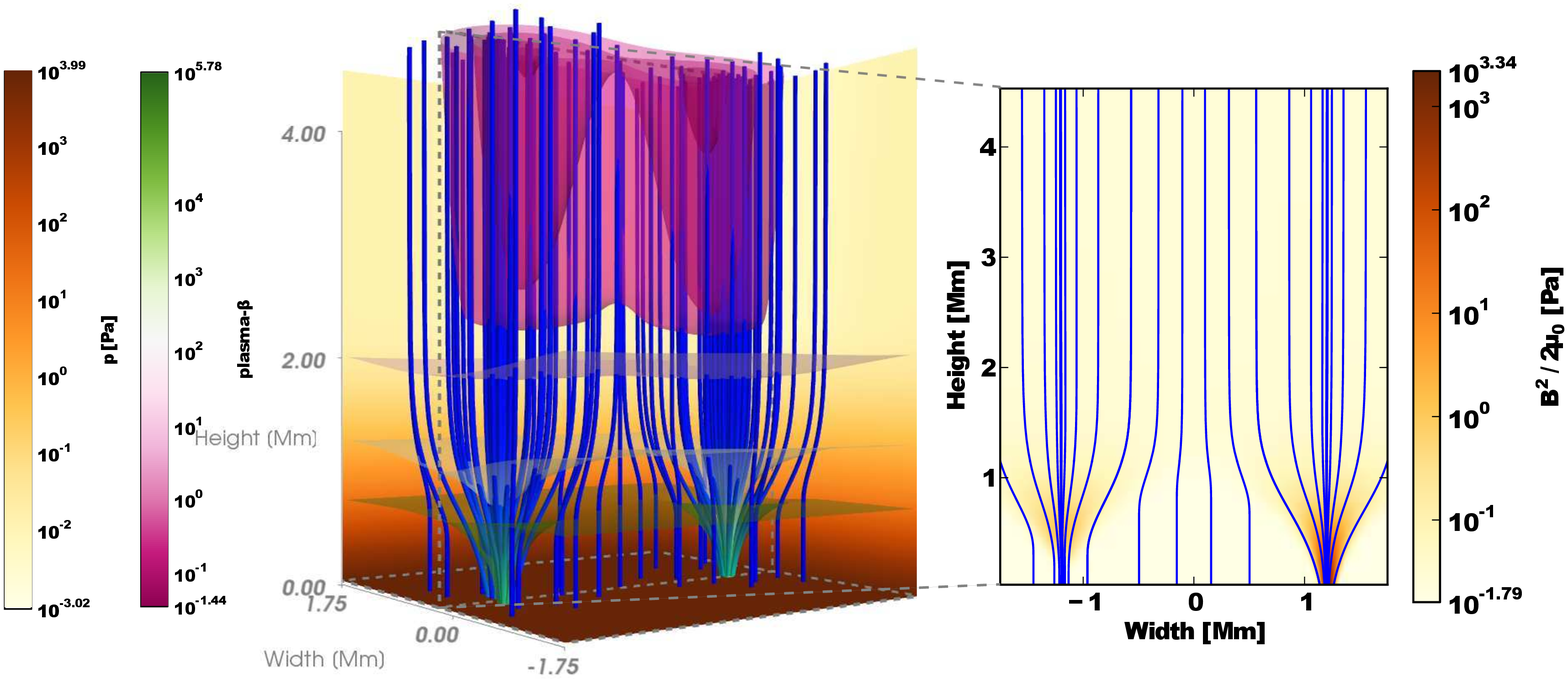}
  \figcaption{
  3D-rendition of two flux tube pairs (left), indicative magnetic
  field lines plotted in blue against a background fill depicting thermal 
  pressure {\apjrev{at the rear and bottom planes}}. 
  Isosurfaces indicate log plasma-$\beta$.
  Right: 2D-slice of magnetic pressure along $y=0$ for the four flux tubes
  (i.e., two pairs) located at axes
  $(x,y)=(1.2,0),(1.2,0),(-1.15,0.15),(-1.25,-0.15)\Mm$, 
  indicative magnetic field lines overplotted in blue.
  The flux tube pair to the right share an identical axis at 
  $(x,y)=(1.2,0)\Mm$, while the axes to the left are slightly separated 
  with respect to both the $x$ and $y$ directions. {\apjrev{Note, the left pair
  have footpoints offset from $y=0$, so the magnetic pressure on this slice
  is highest above the photosphere, where the flux tube pair expand and merge.}}
  \label{fig:2dB}}
  \end{figure*}

  In magnetohydrostatic equilibrium a background atmosphere and magnetic
  field configuration  must satisfy Equation\,\eqref{eq:momb}, which may 
  also be expressed as:
  \begin{equation}\label{eq:delP}
  \nabla p_b 
  + \nabla \frac{|\vect{B}_b|^2}{2\mu_0}
  -\left(\vect{B}_b\cdot\nabla\right)
             \frac{\vect{B}_b}{\mu_0} 
  -\rho_b g\vect{\hat{z}}
  = 
  \vect{0},
  \end{equation}
  where $\vect{\hat{z}}$ is the unit vector and only the global 
  gravitational acceleration directed toward the solar origin is included. 
  The magnetic tension is non-zero due to the curvature of the field lines
  and has opposite sign as it acts as a restoring force to the magnetic
  pressure. 
  We know that the pressure is a scalar quantity, so by taking the curl of 
  $\nabla p$ we obtain
  \begin{equation}\label{eq:curlP}
  \nabla\times\nabla p_b =
  \vect{0}=
  \nabla\times
  \left(
  \rho_b g\vect{\hat{z}} 
  +\frac{\left(\vect{B}_b\cdot\nabla\right)
             \vect{B}_b}{\mu_0} 
  -\nabla \frac{|\vect{B}_b|^2}{2\mu_0}
  \right).
  \end{equation}
  For a single flux tube, this condition is satisfied and a solution for $p_b$ 
  and $\rho_b$ may be obtained similar to that outlined in \GFME.

  We now adopt this approach for a pair of flux tubes whose 
  magnetic configurations are denoted by $\leftidx{^m}{\vect{B}_b}$ and 
  $\leftidx{^n}{\vect{B}_b}$ and  the
  total background magnetic field $\vect{B}_b=\leftidx{^m}{\vect{B}_b}+
  \leftidx{^n}{\vect{B}_b}$.
  Two pairs of such flux tubes,
  with pressure distribution derived as follows,
  are illustrated 
  in Figure\,\ref{fig:2dB}.
  Apart from the magnetic field lines, the plasma-$\beta$ distribution
  also clearly indicates two distinct magnetic structures with
  low plasma-$\beta<1$ in the lower corona and high plasma-$\beta>1$ in the
  photosphere and chromosphere.  
  Note plasma-$\beta\propto {p}/{|\vect{B}|^2}$.
  To make a determination of the necessary pressure distribution for such
  an arrangement, it is useful to decompose the pressure and density into 
  terms that are purely hydrostatic and independent of the 
  magnetic field and terms that represent corrections to balance the 
  effects of the flux tubes.

  Those vertical profiles satisfying the purely hydrostatic 
  background are denoted $p_b\vv$ and $\rho_b\vv$, and are
  derived from the observed vertical profiles
  of \cite{VAL81} and \cite{MTW75} or similar, as described in \GFME.

  Adjustment to the pressure
  distribution required to restore equilibrium due to the inclusion of the 
  magnetic configurations 
  $\leftidx{^m}{\vect{B}_b}$ and $\leftidx{^n}{\vect{B}_b}$
  are denoted by $\leftidx{^m}{p_b\hh}$, $\leftidx{^n}{p_b\hh}$,
  respectively,
  and by $\leftidx{^{mn}}{p_b\hh}$ for their pairwise interaction.
  Hence, $p_b=p_b\vv+\leftidx{^m}{p_b\hh}+\leftidx{^n}{p_b\hh}
  +\leftidx{^{mn}}{p_b\hh}$.
  A corresponding decomposition of the density is also applied, using the same
  superscripts and subscripts.
  Equation\,\eqref{eq:delP}, thus expanded, may then be arranged to yield
  \begin{eqnarray}\label{eq:delPab}
  \nabla (p_b\vv+\leftidx{^m}{p_b\hh}+\leftidx{^n}{p_b\hh}
  +\leftidx{^{mn}}{p_b\hh})=\hspace{2.75cm}&& 
  \\
  \nonumber
  \left[(\leftidx{^m}{\vect{B}_b}
  +\leftidx{^n}{\vect{B}_b})\cdot\nabla\right]
             \frac{(\leftidx{^m}{\vect{B}_b}+\leftidx{^n}{\vect{B}_b})}{\mu_0}
  - \nabla \frac{|\leftidx{^m}{\vect{B}_b}+\leftidx{^n}{\vect{B}_b}|^2}{2\mu_0}&&
  \\
  \nonumber
  +(\rho\vv+\leftidx{^m}{\rho_b\hh}+\leftidx{^n}{\rho_b\hh}
  +\leftidx{^{mn}}{\rho_b\hh}) g\vect{\hat{z}}. &&
  \end{eqnarray}
  In this form, the curl of the right-hand side is not $\vect{0}$ for the 
  self-similar magnetic field, as explained in Appendix\,\ref{sect:curl}.
  In general, there is no valid scalar-field solution to Equation\,\eqref{eq:delPab}
  for $p_b$.
  It makes physical sense that this should be so.
  Pairs of magnetic flux tubes in close proximity will be inclined to 
  attract or repel, depending on their relative polarity, so that
  Equation\,\eqref{eq:delPab} is not in equilibrium.
  Nevertheless, observational evidence exists of multiple flux tubes 
  in 
  relative stability \citep{DRM86,Solanki93,dPTE03}, suggesting the presence of
  some additional   balancing forces.
  These may reside in local anomalies in the neighborhood density and 
  pressure distributions, 
  effects from events at some distance or forces acting at 
  or below the footpoints.
  The nature and source of these forces is complex and beyond the 
  scope of this article, but we conclude that 
  an additional force term is required to satisfy Equation\,\eqref{eq:delPab}.
  \begin{figure*}
  \vspace{-0.75cm}
  \centering
\hspace*{-0.95cm}  \includegraphics[width=0.33\linewidth]{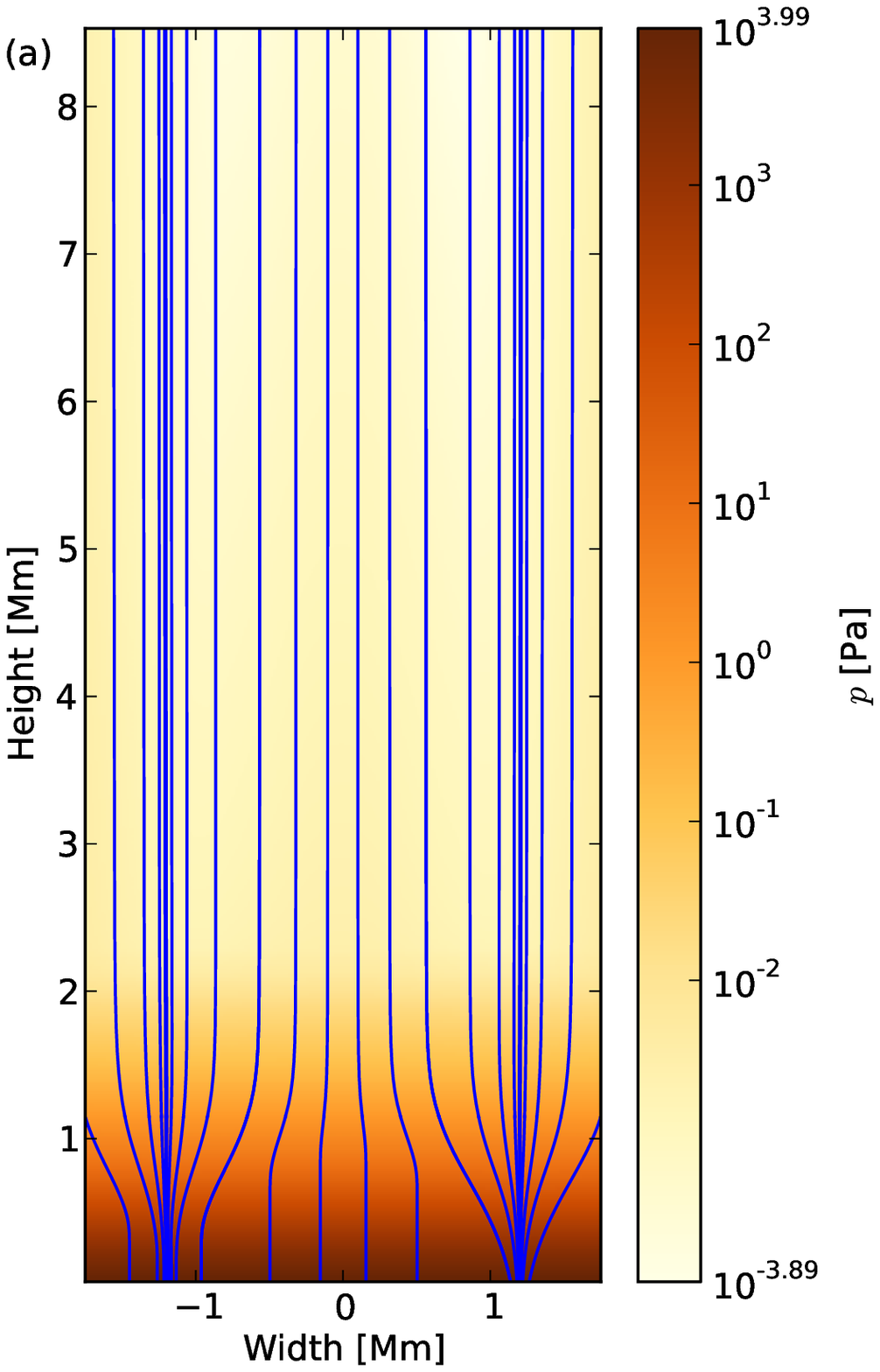}
\hspace*{-0.20cm}  \includegraphics[width=0.33\linewidth]{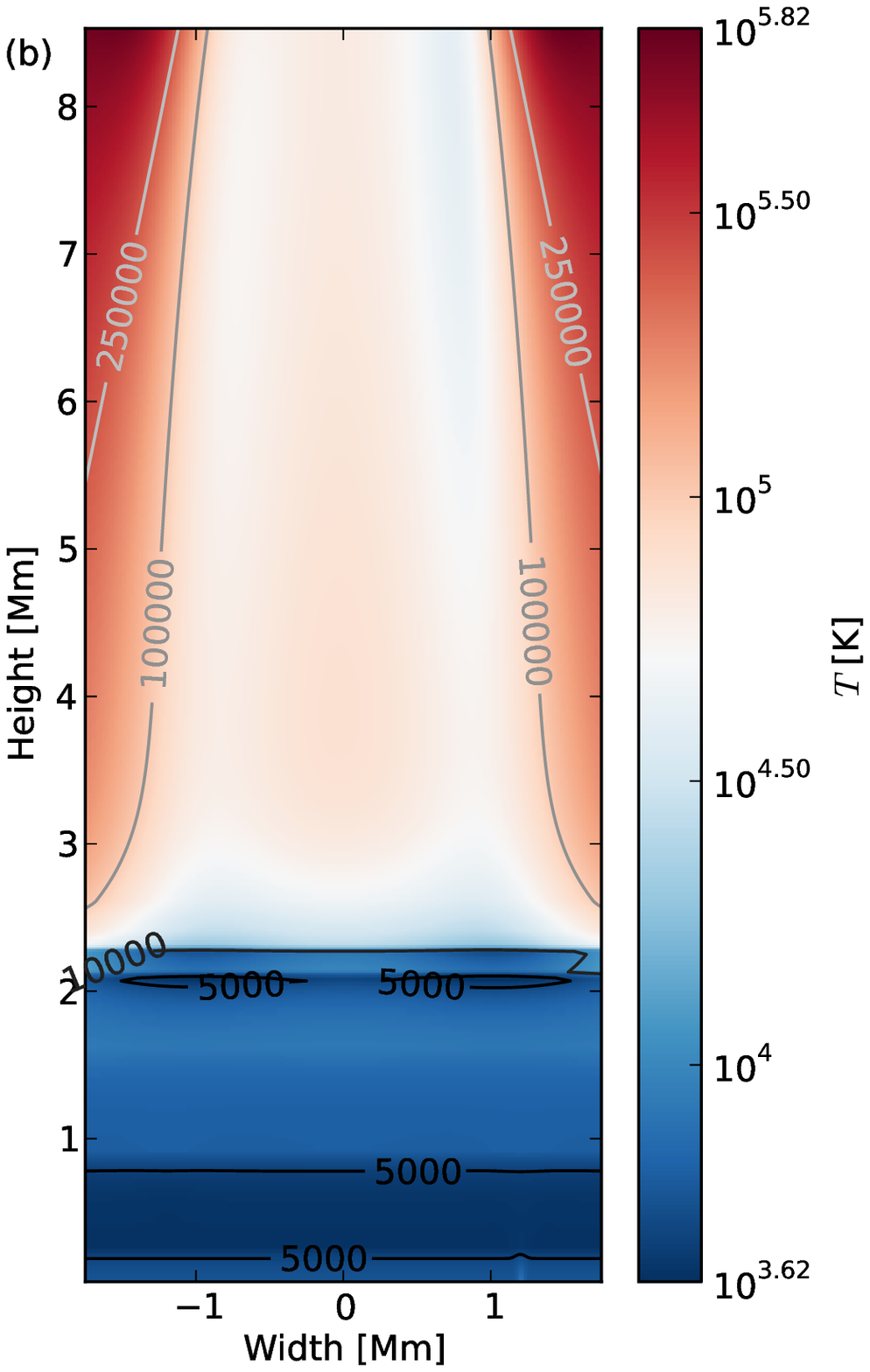}
\hspace*{-0.25cm}  \includegraphics[width=0.33\linewidth]{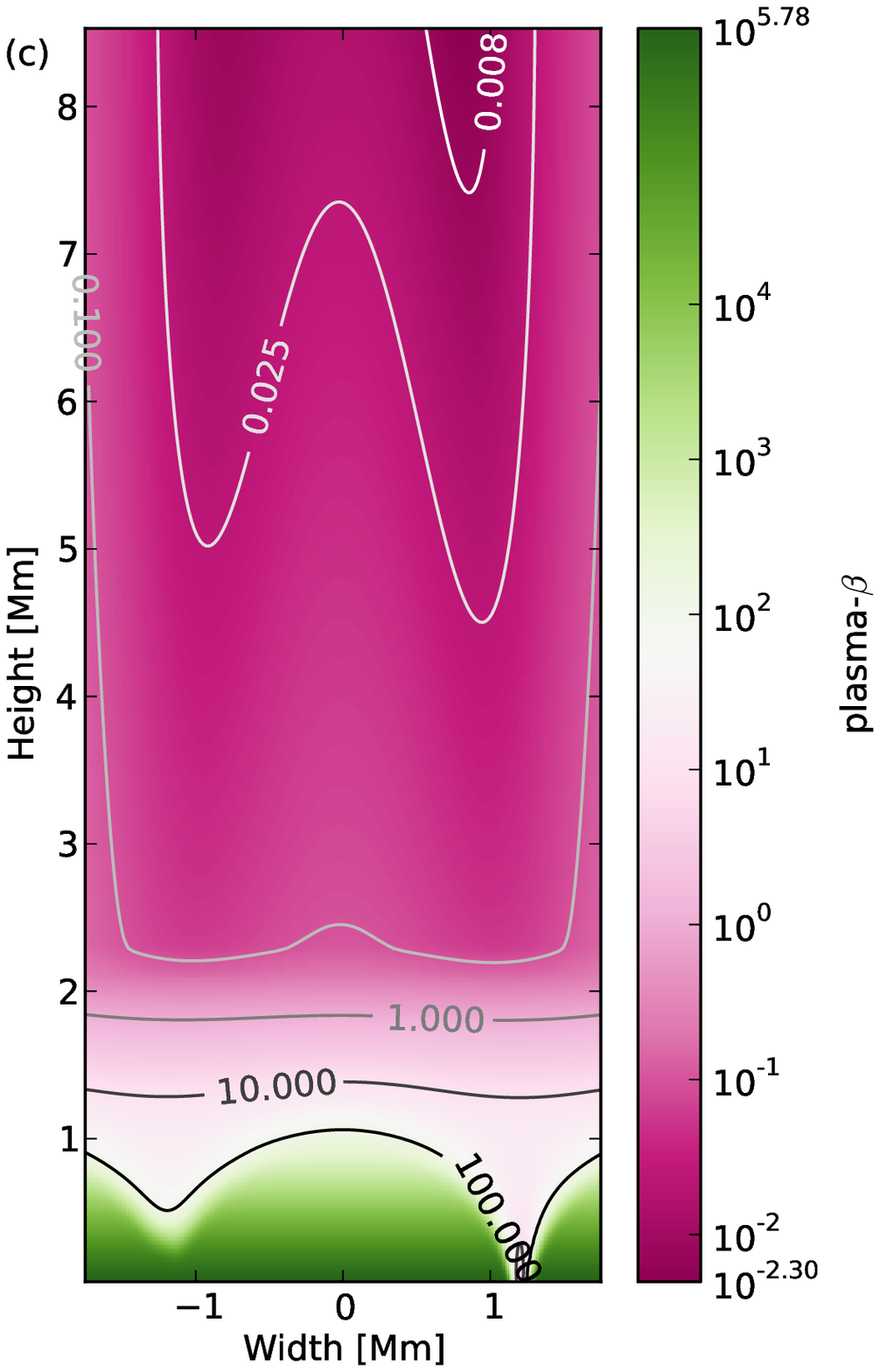}
\hspace*{-0.75cm} \figcaption{
  2D slices along $y=0$ 
  of {{(a)}} thermal pressure
  with indicative magnetic field lines overplotted,
  and {{(b)}} temperature and {{(c)}} plasma beta 
  with line contours overplotted,
  for flux tubes located at axes 
  $(x,y)=(1.2,0),(1.2,0),(-1.15,0.15),(-1.25,-0.15)\Mm$. 
  \label{fig:2dtemp}}
  \end{figure*}

  We can identify the net force acting in Equation\,\eqref{eq:delPab}
  from the non-vanishing terms within
  the right-hand side of Equation\,\eqref{eq:curlP}. 
  First, let us eliminate the terms, for which we already have solutions:
  $$
  \nabla p_b\vv = \rho_b\vv g \vect{\hat{z}}  \quad {\rm and}
  $$and
  $$
  \nabla \leftidx{^m}{p_b\hh}=\leftidx{^m}{\rho_b\hh} g \vect{\hat{z}} 
  - \nabla\frac{|\leftidx{^m}{\vect{B}_b}|^2}{2\mu_0} + (\leftidx{^m}{\vect{B}_b}
    \cdot\nabla)\frac{\leftidx{^m}{\vect{B}_b}}{\mu_0}.
  $$
  The terms $\leftidx{^m}{p_b\hh}$ and $\leftidx{^m}{\rho_b\hh}$ 
  are derived as in \GFME, 
  taking into account the revised definition of the field in 
  Equation\,\eqref{eq:Bxyz}.
  They are detailed in Equations\,\eqref{eq:phfull} and \eqref{eq:rho} of 
  Appendix\,\ref{sect:pbal}, and equivalently $\leftidx{^n}{p_b\hh}$ and 
  $\leftidx{^n}{\rho_b\hh}$.

  This leaves only the interaction terms remaining of Equation\,\eqref{eq:delPab}
  Then, having identified the non-vanishing terms in the right-hand side of 
  Equation\,\eqref{eq:curlP} from amongst these interaction terms,
  as detailed in Appendix\,\ref{sect:curl}, we 
  subtract them to obtain 
  \begin{eqnarray}\label{eq:pbeq}
  \nabla
  \leftidx{^{mn}}{p_b\hh}\,   &=&
  (\leftidx{^m}{\vect{B}_b}
  \cdot\nabla)\frac{\leftidx{^n}{\vect{B}_b}}{\mu_0}
  +(\leftidx{^n}{\vect{B}_b}\cdot\nabla)\frac{\leftidx{^m}{\vect{B}_b}}{\mu_0}
  \nonumber
  \\
  &-&
  \nabla \frac{(\leftidx{^m}{\vect{B}_b}\cdot\leftidx{^n}{\vect{B}_b})}{\mu_0}
  +\leftidx{^{mn}}{\rho_b\hh}\, g\vect{\hat{z}}
  \nonumber
  \\
  &-&
  \left(
   \frac{\leftidx{^m}{B_{bz}}}{\mu_0}\frac{\partial \leftidx{^n}{B_{bx}}}{\partial z}
  +\frac{\leftidx{^n}{B_{bz}}}{\mu_0}\frac{\partial \leftidx{^m}{B_{bx}}}{\partial z}
   \right)
   \vect{\hat{x}}
  \nonumber
  \\
  &-&
  \left(
   \frac{\leftidx{^m}{B_{bz}}}{\mu_0}\frac{\partial \leftidx{^n}{B_{by}}}{\partial z}
  +\frac{\leftidx{^n}{B_{bz}}}{\mu_0}\frac{\partial \leftidx{^m}{B_{by}}}{\partial z}
   \right)
   \vect{\hat{y}},
  \end{eqnarray}
  {{where $\vect{\hat{x}}$ and $\vect{\hat{y}}$
  are unit vectors.}}
  Equation\,\eqref{eq:pbeq} may be solved by considering each vector component in 
  turn: 
  \begin{eqnarray} \label{eq:dpdx}
  \nonumber
  \frac{\partial \leftidx{^{mn}}{p_b\hh}}{\partial x} &=& 
  \frac{\leftidx{^m}{B_{bx}}}{\mu_0}\frac{\partial \leftidx{^n}{B_{bx}}}{\partial x}
  +\frac{\leftidx{^m}{B_{by}}}{\mu_0}\frac{\partial \leftidx{^n}{B_{bx}}}{\partial y}
  +\frac{\leftidx{^n}{B_{bx}}}{\mu_0}\frac{\partial \leftidx{^m}{B_{bx}}}{\partial x}
  \\
  \nonumber
  &+&
  \frac{\leftidx{^n}{B_{by}}}{\mu_0}\frac{\partial \leftidx{^m}{B_{bx}}}{\partial y}
 - \frac{\partial }{\partial x}
  \frac{\left(\leftidx{^m}{\vect{B}_b}\cdot\leftidx{^n}{\vect{B}_b}
  \right)}{\mu_0},
  \end{eqnarray} 
  \begin{eqnarray} \label{eq:dpdy}
  \nonumber
  \frac{\partial \leftidx{^{mn}}{p_b\hh}}{\partial y}\, &=& 
  \frac{\leftidx{^m}{B_{bx}}}{\mu_0}\frac{\partial \leftidx{^n}{B_{by}}}{\partial x}
  +\frac{\leftidx{^m}{B_{by}}}{\mu_0}\frac{\partial \leftidx{^n}{B_{by}}}{\partial y}
  +\frac{\leftidx{^n}{B_{bx}}}{\mu_0}\frac{\partial \leftidx{^m}{B_{by}}}{\partial x}
  \\
  &+&
  \frac{\leftidx{^n}{B_{by}}}{\mu_0}\frac{\partial \leftidx{^m}{B_{by}}}{\partial y}
-  \frac{\partial }{\partial y}
  \frac{\left(\leftidx{^m}{\vect{B}_b}\cdot\leftidx{^n}{\vect{B}_b}
  \right)}{\mu_0}
  \end{eqnarray} 
  and
  \begin{eqnarray} \label{eq:dpdz}
  \nonumber
  \frac{\partial \leftidx{^{mn}}{p_b\hh}}{\partial z} &=& 
  \frac{\leftidx{^m}{B_{bx}}}{\mu_0}\frac{\partial \leftidx{^n}{B_{bz}}}{\partial x}
  +\frac{\leftidx{^m}{B_{by}}}{\mu_0}\frac{\partial \leftidx{^n}{B_{bz}}}{\partial y}
  +\frac{\leftidx{^m}{B_{bz}}}{\mu_0}\frac{\partial \leftidx{^n}{B_{bz}}}{\partial z}
  \\
  \nonumber
  &+&
  \frac{\leftidx{^n}{B_{bx}}}{\mu_0}\frac{\partial \leftidx{^m}{B_{bz}}}{\partial x}
  +\frac{\leftidx{^n}{B_{by}}}{\mu_0}\frac{\partial \leftidx{^m}{B_{bz}}}{\partial y}
  +\frac{\leftidx{^n}{B_{bz}}}{\mu_0}\frac{\partial \leftidx{^m}{B_{bz}}}{\partial z}
  \\
  &+&
  \leftidx{^{mn}}{\rho_b\hh} g 
  -\frac{\partial }{\partial z}
  \frac{\left(\leftidx{^m}{\vect{B}_b}\cdot\leftidx{^n}{\vect{B}_b}
  \right)}{\mu_0}.
  \end{eqnarray} 
  The solution for $\leftidx{^{mn}}{p_b\hh}$ 
  is specified by Equation\,\eqref{eq:apx} and $\leftidx{^{mn}}{\rho_b\hh}$ by 
  Equation\,\eqref{eq:rhoijg} and detailed in Appendix\,\ref{sect:pbal}.
  We can now specify the background pressure and density profiles 
  in the 3D space as
  $p_b = p_b\vv+\leftidx{^m}{p_b\hh}+\leftidx{^n}{p_b\hh}
  +\leftidx{^{mn}}{p_b\hh}$ and 
  $\rho_b=\rho_b\vv+\leftidx{^m}{\rho_b\hh}+\leftidx{^n}{\rho_b\hh}
  +\leftidx{^{mn}}{\rho_b\hh}$.
  Note that numeric values of the last three terms in the expressions for $p_b$ 
  and $\rho_b$ can be negative. 
  A minimal constraint on the choice of parameters for the magnetic field 
  configuration must be that both sums are sufficiently positive to guarantee 
  that $p_b+\tilde{p}>0$ and $\rho_b+\tilde\rho>0$ everywhere while the 
  system is being perturbed.

  Figure\,~\ref{fig:2dtemp} shows vertical slices of the pressure and 
  temperature
  for two pairs of flux tubes as well as the resulting plasma-$\beta$.
  Two flux tubes exactly co-located on the right form a slightly stronger 
  magnetic structure, while on the left, two identical flux tubes are
  slightly separated to form a weaker configuration.
  The delineation into two distinct combinations is most evident in the 
  temperature and plasma-$\beta$ distributions.
  The radial symmetry is broken and there is a clear opportunity to investigate
  the interaction between the flux tubes, although they are almost identical.

  For more challenging configurations, we require more flux tubes with
  irregular spacing.
  First, we need to consider the consequences of this configuration for the 
  MHD equations.

\subsection{Consequences for the MHD Equations}\label{sect:change}

  In Section \ref{subsect:initHD}, we have derived profiles for
  the background pressure and plasma density.
  Suppose that $p_b$ and $\rho_b$ are now thus defined for the magnetic flux 
  tube
  pair $\leftidx{^1}{\vect{B}_b}$ and $\leftidx{^2}{\vect{B}_b}$.
  If we subtract Equation\,\eqref{eq:momb} from the unperturbed 
  Equation\,\eqref{eq:mom},
  as presented in Section\,\ref{subsect:MHD}, we then obtain
  \begin{eqnarray}  
    \label{eq:diseq}
    \rho_b\frac{\partial u_i}{\partial t} &=&
    \frac{\partial}{\partial x_i}
    \left(
    p_b + \frac{B_{bj}B_{bj}}{2\mu_0}
    \right)
    -\frac{\partial}{\partial x_j}
    \left(
    \frac{B_{bi}B_{bj}}{\mu_0}
    \right)- \rho_b g_i 
    \nonumber 
    \\
    &=&
    -(\delta_{1i} + \delta_{2i})
    \left[
    \frac{\leftidx{^1}{B_{b3}}}{\mu_0}
    \frac{\partial \leftidx{^2}{B_{bi}}}{\partial x_3}
    +
    \frac{\leftidx{^2}{B_{b3}}}{\mu_0}
    \frac{\partial \leftidx{^1}{B_{bi}}}{\partial x_3}
    \right],
  \end{eqnarray}  
  where $(x_1,x_2,x_3) = (x,y,z)$ and $\delta_{ij}$ is 
  the Kronecker delta and the latter equality arises
  from the  supplementary terms applying in Equation\,\eqref{eq:pbeq}.
  That is, the system is out of equilibrium because the equality of 
  Equation\,\eqref{eq:momb} is no longer valid for the multi-flux tube
  configuration.
  In this form the advective term from Equation\,\eqref{eq:momp} would also
  contribute, as $u_i\neq0_i$ for the background state. 
  However, by restoring equilibrium as follows, this term reverts to zero.
  \begin{center}
  \includegraphics[width=0.8\linewidth]{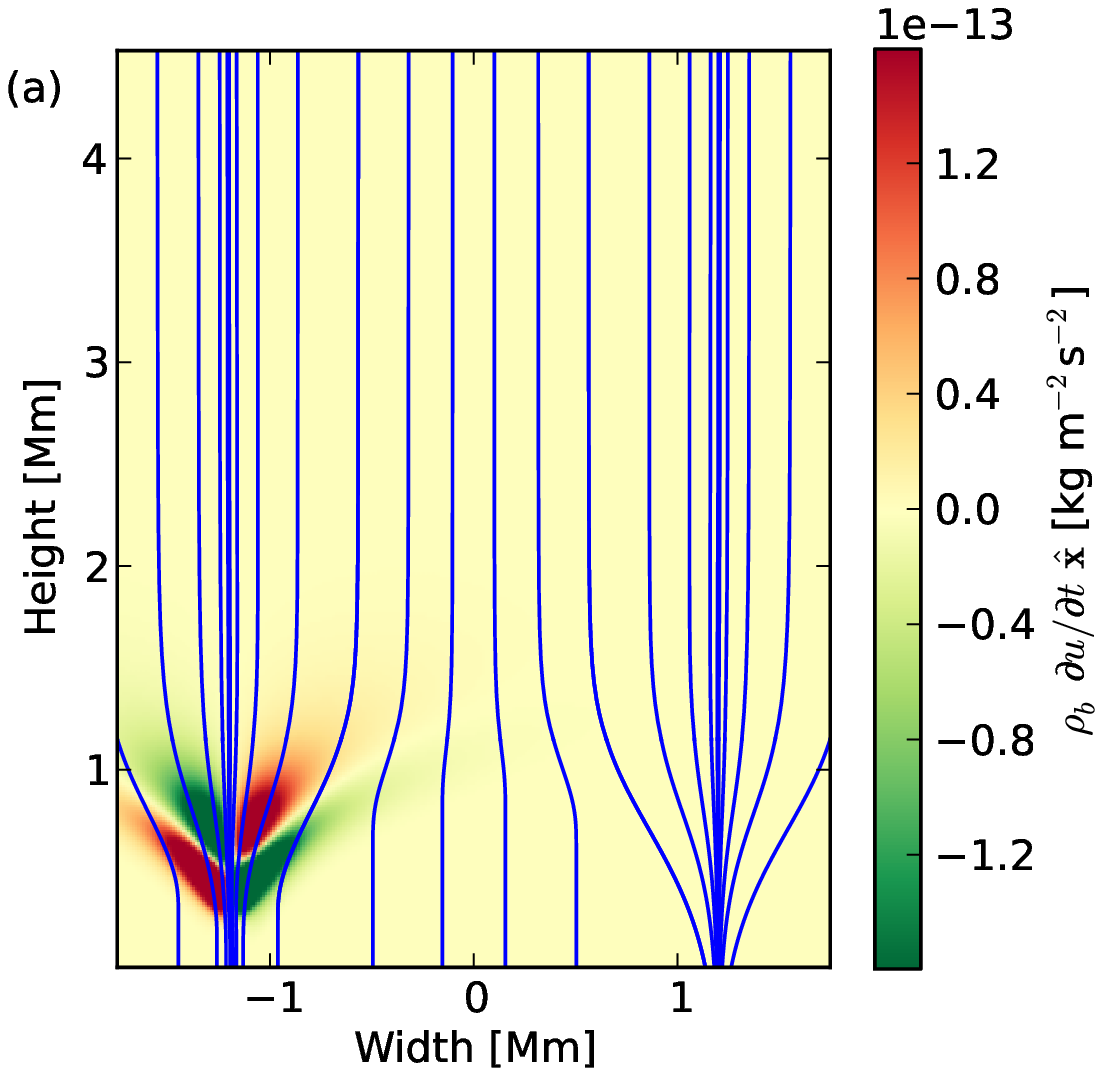}\\
  \includegraphics[width=0.8\linewidth]{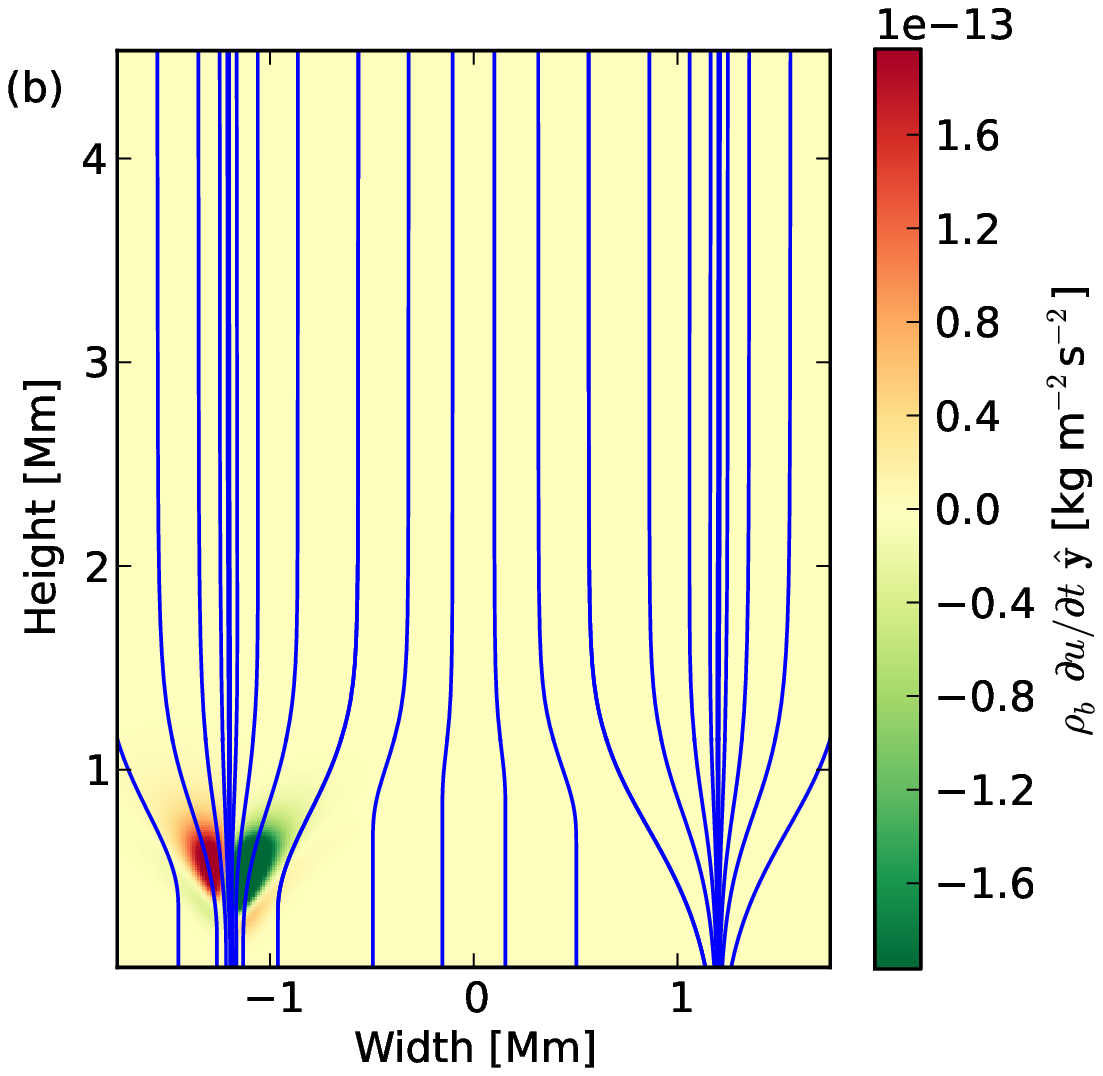}
  \figcaption{
  2D slices along $y=0$ 
  of forces applying in Equation\,\eqref{eq:Fbalsum} 
  for flux tubes at axes
  $(x,y)=(1.2,0),(1.2,0),(-1.15,0.15),(-1.25,-0.15)\Mm$
  of the $\vect{\hat{x}}$-component {{(a)}}  
  and $\vect{\hat{y}}$-component {{(b)}}. 
  Indicative magnetic field lines are overplotted in blue.
  \label{fig:Fbal}}
  \end{center}

  The ${{x}}$-component, Figure\,\ref{fig:Fbal} panel (a), and the 
  ${{y}}$-component, panel (b), of the net forces from
  Equation\,\eqref{eq:diseq} applying along $y=0$ for two pairs of flux tubes
  are illustrated. 
  The pair of flux tubes on the right share an identical axis at 
  $(x,y)=(1.2,0)\Mm$, so the
  forces applying between them are zero on $y=0$, whereas on the left the
  flux tube
  axes are slightly separated with respect to both the $x$ and $y$ directions.
  These forces are most evident in the chromosphere, where the field curvature
  is strongest.
  Alternating azimuthal trajectories of the forces with radius and height
  suggest that mutual torsional oscillations would result in the absence of any 
  balancing forces. 
  There are also forces acting between both pairs, but these are negligible
  near the footpoints because of the large axial separation and in the corona
  because of the weakness of the tension forces. 

  This magnetic field, with two flux tubes may be expressed as 
  $  \vect{B} = \leftidx{^1}{\vect{B}_b} + \leftidx{^2}{\vect{B}_b} + \tilde{\vect{B}}$.
  Thus,
$    |\vect{B}|^2 = |\leftidx{^1}{\vect{B}_b}|^2 + |\leftidx{^2}{\vect{B}_b}|^2
                 + |\tilde{\vect{B}}|^2 
                 + 2\leftidx{^1}{\vect{B}_b}\cdot\tilde{\vect{B}}
                 + 2\leftidx{^2}{\vect{B}_b}\cdot\tilde{\vect{B}}
                 + 2\leftidx{^1}{\vect{B}_b}\cdot\leftidx{^2}{\vect{B}_b}$,
  with an equivalent pairwise decomposition also applying for the tension 
  force $(\vect{B}\cdot\nabla)\vect{B}$. 
  Hence, terms pertaining to interactions between a pair of background flux 
  tubes $\leftidx{^1}{\vect{B}_b}$ and $\leftidx{^2}{\vect{B}_b}$ exclude 
  the perturbations $\tilde{\vect{B}}$.
  
  If we extend this to a network of $N$ magnetic flux tubes with the 
  configuration  $\leftidx{^m}{\vect{B}_b}$ as 
  defined by Equation\,\eqref{eq:Bxyz}, but free to differ by axial location 
  $(\leftidx{^m}x,\leftidx{^m}y) $, then the magnetic field can be expressed by the sum
  \[
    \vect{B} = \vect{B}_b + \tilde{\vect{B}} 
             = \left(
                 \sum_{m=1}^{N}\leftidx{^m}{\vect{B}_b}
               \right) +\tilde{\vect{B}},
  \]
  and the restriction to pairwise interactions between each $m^{\rm th}$ 
  flux tube, its neighbors and
  the perturbed magnetic field still applies.
  The latter equality from Equation\,\eqref{eq:diseq} can be generalized for
  $N$ flux tubes to 
  \begin{eqnarray}\label{eq:Fbalsum}
     \rho_b\frac{\partial\vect{u}}{\partial t} =
     \vect{F}_{\rm bal} = &-&\hspace{-0.5cm} 
\sum_{m,n=1|m\neq n}^N\hspace{-0.2cm} 
     \frac{\leftidx{^n}{B_{bz}}}{\mu_0}\frac{\partial \leftidx{^m}{B_{bx}}}{\partial z}\vect{\hat{x}}
     \\ \nonumber&-&\hspace{-0.5cm}
\sum_{m,n=1|m\neq n}^N\hspace{-0.2cm} 
     \frac{\leftidx{^n}{B_{bz}}}{\mu_0}\frac{\partial \leftidx{^m}{B_{by}}}{\partial z}\vect{\hat{y}}
     +0\vect{\hat{z}},
  \end{eqnarray}
  in which the explicit expression for each $mn$ pairing is 
  given in Equation\,\eqref{eq:Fij} of Appendix\,\ref{sect:curl}.
  To satisfy the equality with $\rho\vect{g}$, Equation\,\eqref{eq:momb} must be revised to
  \begin{equation}
    \label{eq:pbeqb}
    \frac{\partial}{\partial x_i}
    \left(
    p_b + \frac{B_{bj}B_{bj}}{2\mu_0}
    \right)
    -\frac{\partial}{\partial x_j}
    \left(
    \frac{B_{bi}B_{bj}}{\mu_0}
    \right) 
    - F_{{\rm bal}_i}
     = \rho_b g_i.
  \end{equation}
  Scalar multiplication of Equation\,\eqref{eq:Fbalsum} with $\vect{u}$ then yields
  \begin{equation}
    \label{eq:Ebal}
     \vect{F}_{\rm bal}\cdot\vect{u} = - u_x\hspace{-0.5cm}\sum_{m,n=1|m\neq n}^N \hspace{-0.2cm}
     \frac{\leftidx{^n}{B_{bz}}}{\mu_0}\frac{\partial \leftidx{^m}{B_{bx}}}{\partial z}
     - u_y\hspace{-0.5cm}\sum_{m,n=1|m\neq n}^N\hspace{-0.2cm}
     \frac{\leftidx{^n}{B_{bz}}}{\mu_0}\frac{\partial \leftidx{^m}{B_{by}}}{\partial z}
  \end{equation}
  and Equation\,\eqref{eq:energyb} must also be revised as
  \begin{equation}
    \label{eq:energyr}
    u_i\frac{\partial}{\partial x_i}
    \left(
    p_b + \frac{B_{bj}B_{bj}}{2\mu_0}
    \right)
    -u_i\frac{\partial}{\partial x_j}
    \left(
    \frac{B_{bi}B_{bj}}{\mu_0}
    \right)
    -F_{{\rm bal}i}u_i
     = \rho_b g_iu_i.
  \end{equation}
  Now, subtracting Equation\,\eqref{eq:pbeqb} from Equation\,\eqref{eq:mom} and
  Equation\,\eqref{eq:energyr} from Equation\,\eqref{eq:energy}, we obtain the revised
  MHD equations for the perturbed momentum and energy 
  \begin{eqnarray}
    \label{eq:momf}
    \frac{\partial [(\rho_b+\tilde\rho) u_i]}{\partial t}
    +\frac{\partial
    }{\partial x_j}
    \left[
    (\rho_b+\tilde\rho) u_iu_j
    + \tilde p_T 
    \right]
    &&\\\nonumber
    -\frac{\partial}{\partial x_j}
    \left[
    \frac{\tilde{B_i}B_{bj}+B_{bi}\tilde{B_j}+\tilde{B_i}\tilde{B_j}}{\mu_0}
    \right]
    +F_{{\rm bal}i}
    & =& \tilde\rho g_i,\\
    \label{eq:energyf}
    \frac{\partial \tilde{e}}{\partial t}
    +\frac{\partial}{\partial x_j}
    \left[
    (e_b+\tilde{e}) u_j 
    -\frac{\tilde{B_i}\tilde{B_j}}{\mu_0}u_i
    +\tilde p_T u_j 
    \right] &&\\
    -\frac{\partial}{\partial x_j}
    \left[
    \frac{\tilde{B_i}B_{bj}+B_{bi}\tilde{B_j}}{\mu_0}u_i
    \right]
    &&\nonumber\\
    +p_{bT}\frac{\partial u_j}{\partial x_j} 
    - \frac{B_{bj}B_{bi}}{\mu_0} \frac{\partial u_i}{\partial x_j}
    +F_{{\rm bal}i}u_i
     & =&
    \tilde\rho g_i u_i.\nonumber
  \end{eqnarray}
  With the addition of $\vect{F}_{\rm bal}$ in Equation\,\eqref{eq:momf} the 
  unperturbed system has 
     \[
     \rho_b\frac{\partial\vect{u}}{\partial t} = \vect{0}.
     \]
  Note also that this does not affect any terms depending on the perturbations 
  and is independent of changes to the perturbed system,
  so it remains constant over time and the background is in equilibrium.

  The corresponding term $\vect{F}_{\rm bal}\cdot\vect{u}$ in
  Equation\,\eqref{eq:energyf} is zero in the steady state, but is apparently 
  subject to amplification by horizontal components of the velocity field.
  However, an equal and opposite effect is present due to the subtraction of
  the other terms in Equation\,\eqref{eq:energyr}, so these combine to 
  result in zero net energy effect.

\subsection{Inhomogeneous Multiple Flux Tubes}\label{sect:multi}
  
  As outlined in Section\,\ref{sect:change},
  when adding multiple flux tubes, the background magnetic pressure gradient
  and tension force are fully specified by the sum of single and pairwise
  interactions between each flux tube. 
  Thus, given a magnetic field comprising $N$ magnetic flux tubes,
  \begin{equation}\label{eq:sumB}
    \vect{B}_b = \sum_{m=1}^N\leftidx{^m}{\vect{B}_b},
  \end{equation} 
  \begin{equation}\label{eq:sump}
    p_b=p_b\vv+\sum_{m=1}^N\leftidx{^m}{p_b\hh} 
       +{\sum_{m,n=1|n>m}^{N}} \leftidx{^{mn}}{p_b\hh} 
  \end{equation} and
  \begin{equation}\label{eq:sumr}
    \rho_b=\rho_b\vv+\sum_{m=1}^N \leftidx{^m}{\rho_b\hh} +{\sum_{m,n=1|n>m}^{N}}
    \leftidx{^{mn}}{\rho_b\hh}, 
  \end{equation} 
  in which $\leftidx{^{mn}}{p_b\hh}$ and $\leftidx{^{mn}}{\rho_b\hh}$ 
  represent the action of $\leftidx{^m}{\vect{B}_b}$
  on $\leftidx{^n}{\vect{B}_b}$ \emph{and} vice versa. 
  Hence, the inequality under the summation is
  required for this quantity to be counted only once for each pair of flux
  tubes.

  The time-independent momentum equation describing the background equilibrium
  is then 
  \begin{equation}
    \label{eq:momm}
    \rho_b g\vect{\hat{z}}-\nabla p_b - \nabla\frac{|\vect{B}_b|^2}{2\mu_0}
    + \frac{(\vect{B}_b\cdot\nabla)\vect{B}_b}{\mu_0}
    + \vect{F}_{\rm bal} = \vect{0},
  \end{equation}
  where $\vect{F}_{\rm bal}$ is as specified by Equation\,\eqref{eq:Fbalsum}. 
  Note that this solution will yield a different equilibrium configuration to 
  the solution of Equation\,\eqref{eq:delP} (Section\,\ref{subsect:initHD}),
  valid for a single flux tube.
  Consider some single flux tube with $\vect{F}_{\rm bal}\equiv\vect{0}$.
  Let us construct an identical flux tube by combining some field 
  configurations with a common axis of the form $\leftidx{^m}{\vect{B}_b}$,
  then $\vect{F}_{\rm bal}$ will be non-zero. 
  Therefore, the solution of Equation\,\eqref{eq:delP} for the single flux tube
  will obtain different distributions for pressure and density to a solution
  of Equation\,\eqref{eq:momm} for an identical magnetic field, with the former 
  equilibrium satisfying Equation\,\eqref{eq:momp} and the latter 
  Equation\,\eqref{eq:momf}.
 
  We have devised a background magnetic field construction by
  the summation of multiple locally defined field configurations in 
  magnetohydrostatic equilibrium with the stratified atmosphere, spanning the
  transition between the solar photosphere and lower solar corona.
  Let us consider some opportunities presented by this arrangement.

  \begin{center}
 \hspace*{0cm} \includegraphics[width=0.95\linewidth]{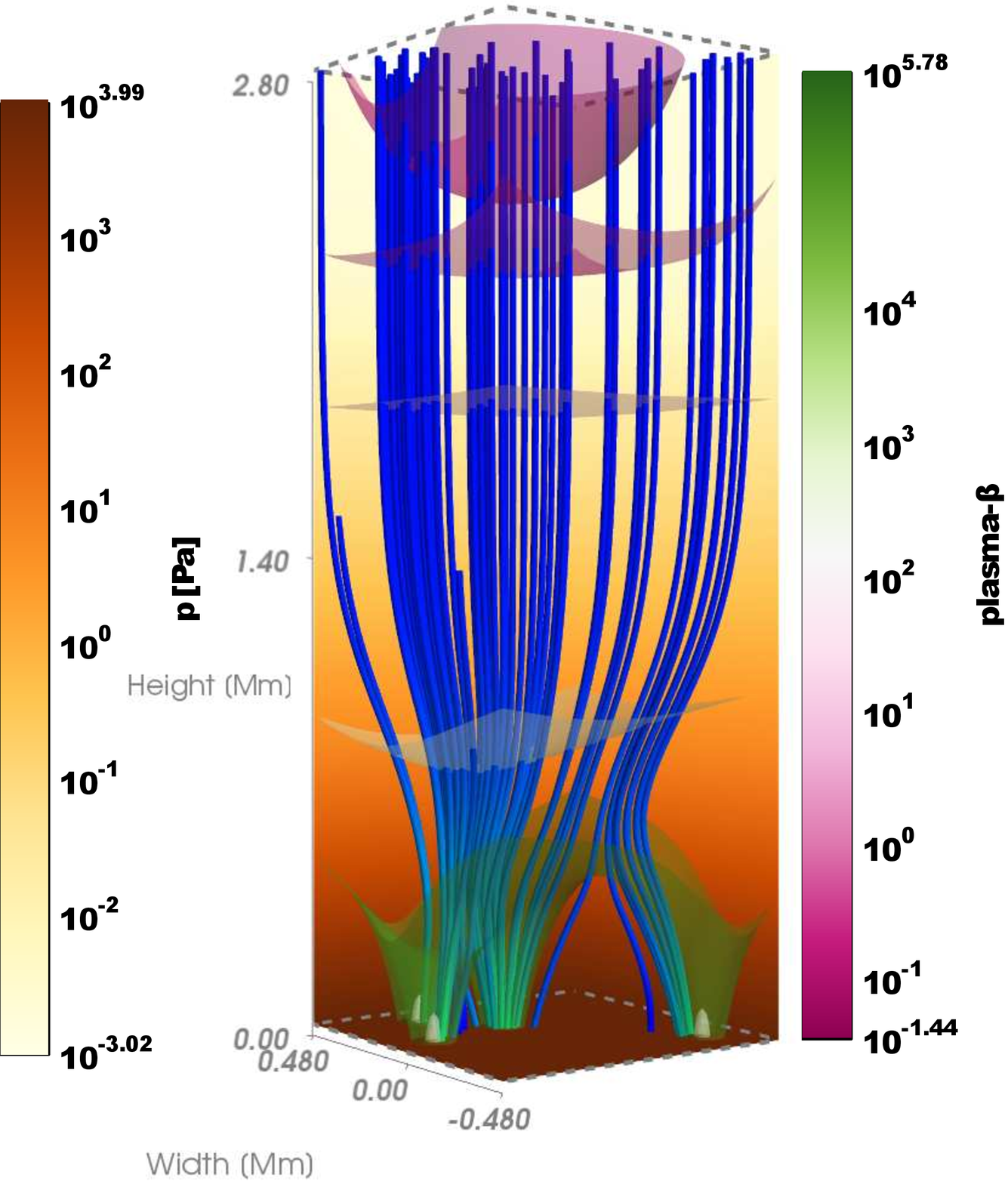}
  \figcaption{
   Magnetic field comprising flux tube 
   sources with independent axes
   located at $(x,y) = (0.34, 0.20)$, $(-0.31,-0.34)$, $(0.07,0.33)$
    and $(0.14,0.04)\Mm$.
   Sample magnetic field lines are overplotted in blue and isosurfaces
   indicate the variation in plasma-$\beta$.
   Background fill shows thermal pressure {\apjrev{on those planes}}.
  \label{fig:single}}
  \end{center}

  On scales below the minimum observable resolution, the fine structure of
  the magnetic field can add to the complexity and dynamics of a magnetic 
  flux tube.
  We combine four independent magnetic sources clustered within a photospheric
  surface element $1"\times1"\approx725\km \times 725\km$.
  This corresponds to the maximum resolution for the magnetic field 
  observations of, for example, the Helioseismic and Magnetic Imager of
  the Solar Dynamics Observatory \citep{HMI07}.
  Hence, the fine structure of a magnetic field configuration below this 
  resolution would be treated as a single flux tube, but may well be, and most
  likely is, the combination of an irregular magnetic field network.

  We thus construct a non-axisymmetric background
  magnetic field, which, in the corona, forms a single identifiable 
  structure, but in the chromosphere has significant complexity. 
  Although the field lines merge in the corona, they retain complexity in the 
  form of pressure, density, and plasma-$\beta$ fluctuations.
  An example of such an arrangement is illustrated in Figure\,\ref{fig:single}.
  Perturbations to this steady background will be subject to nonlinear 
  effects in the horizontal direction, due to the irregular field strength, and
  also in the vertical direction, due to the pressure gradient and the
  transition from the high to low plasma-$\beta$ regime.
 
  \begin{figure*}
  \centering
  \hspace*{-1cm}\includegraphics[width=1.08\linewidth]{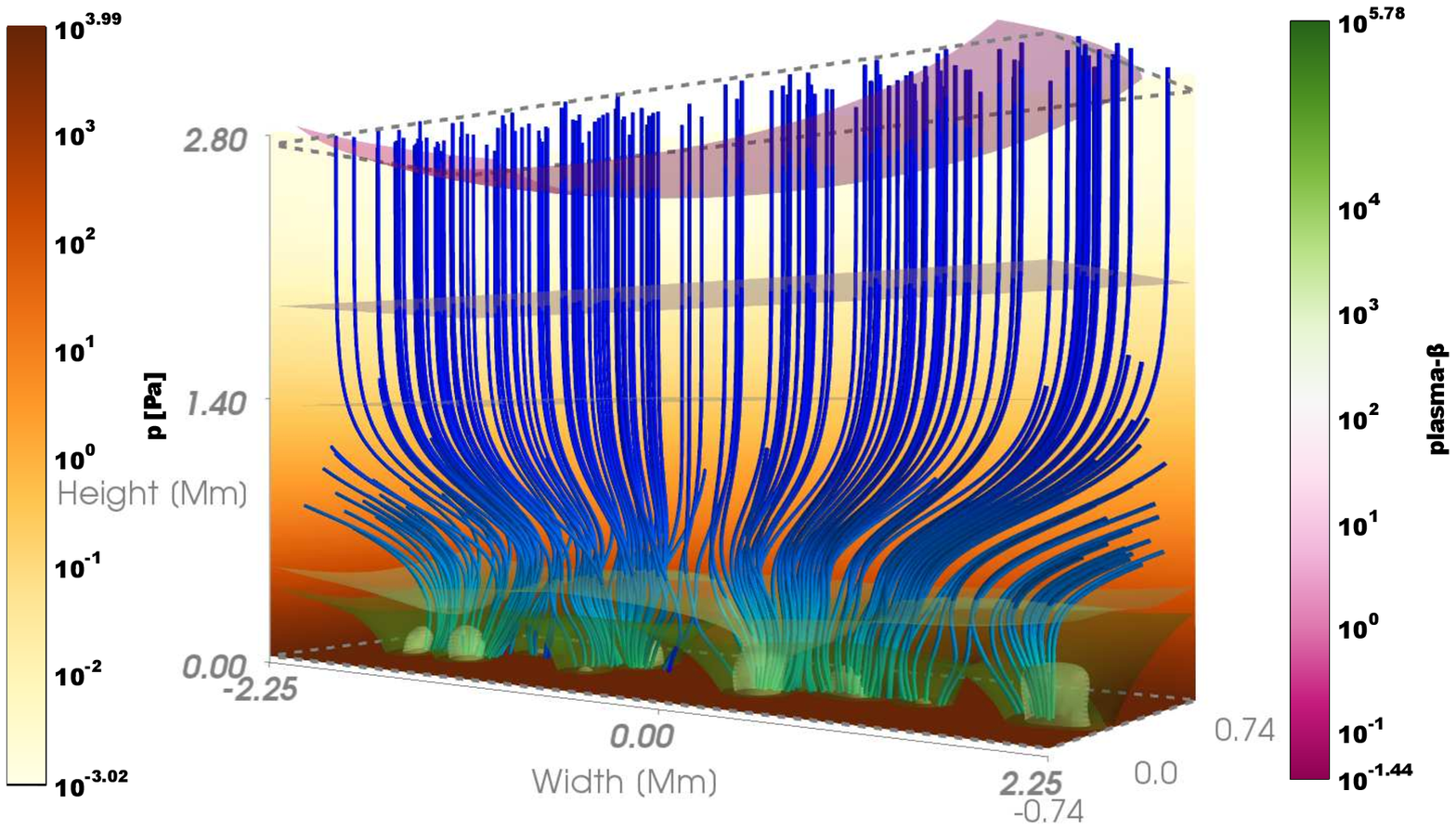}
  \figcaption{
  Example of magnetic field configuration modeling granular lanes within the
  photosphere. 
  Multiple flux tubes emerging through the chromosphere along a narrow lane
  merge in the corona to form a magnetic canopy.
  \label{fig:3dplot}}
  \end{figure*}

  For the same field configuration a 2D horizontal slice at $z=0.5\Mm$ of the
  steady background thermal pressure profile is shown in Figure\,\ref{fig:pxy}.
  The deviations in the plasma pressure are small compared to the vertical 
  differential.
  Overplotted in blue are some magnetic field lines.
  As might be expected, field lines emanate from the flux tube axes, 
  indicated by the light (low {\apjrev{pressure}} regions).
  {\apjrev{Above the three footpoint axes located at the photosphere}} in the
  positive $x,y$ quadrant the pressure has already merged into a single 
  depression. 
  \begin{center}
  \hspace*{0.1cm}\includegraphics[width=1.1\linewidth]{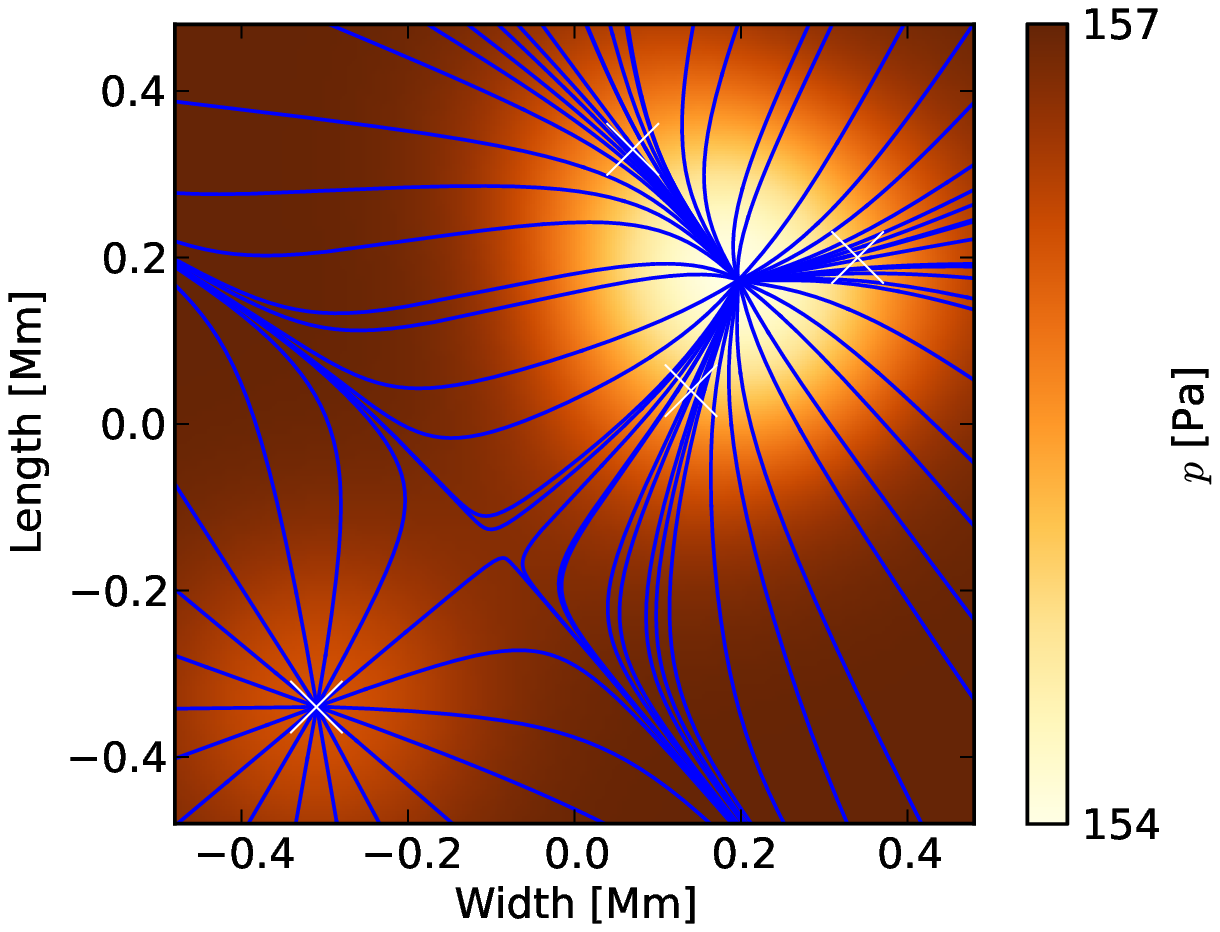}
  \figcaption{
  2D horizontal slice of thermal pressure at $z=0.5\Mm$
  for the magnetic field comprising flux tube sources with 
  {\apjrev{photospheric footpoint axes independently}}
  located at $(x,y) = (0.34, 0.20)$, $(-0.31,-0.34)$, $(0.07,0.33)$
  and $(0.14,0.04)\Mm${\apjrev{, as identified in the figure by white crosses}}.
  {{Due to radial expansion, by $z\geq0.5\Mm$ the three flux tubes in the
  positive quadrant have merged.}}
  Indicative horizontal magnetic field lines are overplotted in blue,
  {{diverted where the flux tubes intersect.}} 
  \label{fig:pxy}}
  \end{center}
  By the height of the TR, the smaller depression  
  {\apjrev{within the flux tube anchored to the photosphere at 
  }} $(x,y) = (-0.31,-0.34)\Mm$ also merges with the other three to form a 
  single, non-uniform low pressure core inside the composite magnetic flux tube.
  In the plane field lines are not purely radial, with
  azimuthal trajectories appearing due to the influence of neighboring flux 
  tubes. 
  Between the axis at $(-0.31,-0.34)\Mm$ and the other three axes field lines 
  with opposite polarity appear to meet, and in the regions between the three
  positive
  axes there are high concentrations of field lines as they merge with each
  other.
  In 3D, these lines do not meet due to the vertical component of the
  field.
  However magneto-acoustic waves along these converging field lines 
  may interfere with each other near these intersections.

  In ideal MHD there is no mechanism for reconnection. 
  For numerical stability, however, simulations require a minimum level of
  numerical diffusion.
  Such diffusion will be strongest in regions where the field lines converge,
  resulting in topological changes to the field configuration analogous
  to magnetic reconnection.

  In addition to providing an 
  interesting structure for a single flux tube as in the
  preceding example, it is possible to construct networks of flux tubes on 
  larger scales.
  Figure\,\ref{fig:3dplot} illustrates a 3D rendition of the magnetic field 
  resembling a granular lane.
  This could be extended to form a ring or other network of flux tubes.
  With a sufficiently large numerical domain, the horizontal interactions in the
  corona between
  flux tubes and networks of flux tubes can be explored.
  Even in the corona, on this scale the field can exhibit much more 
  anisotropy.

  \begin{figure*}
  \centering
 \hspace*{-1cm} \includegraphics[width=1.05\linewidth]{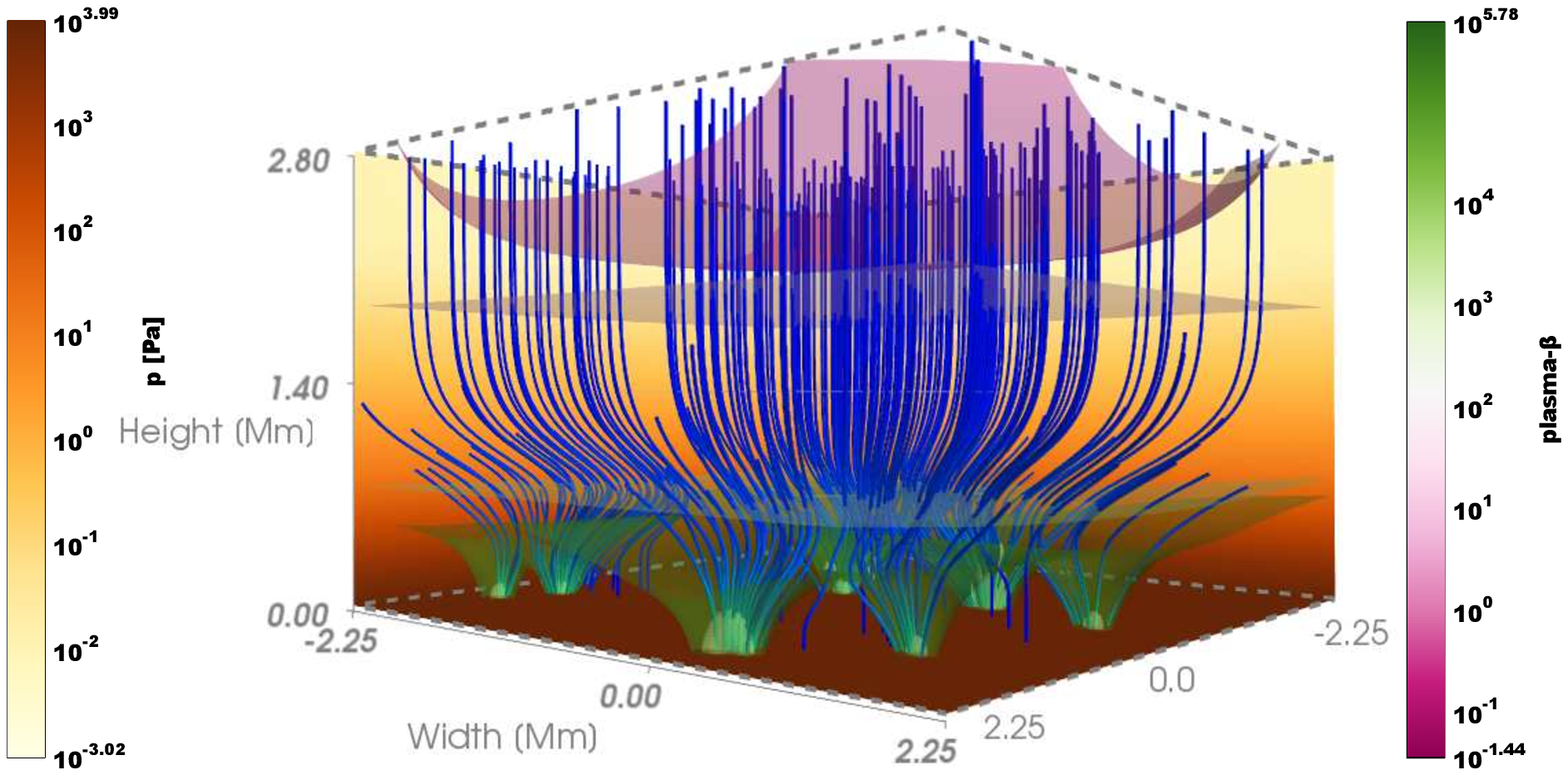}
  \figcaption{
  Example of magnetic field configuration modeling an active region with 
  multiple magnetic bright points and emerging flux tubes.
  The strong compact individual magnetic flux tubes in the chromosphere spread 
  with height and merge to fill the corona.
  \label{fig:parker}}
  \end{figure*}

  In Figure\,\ref{fig:parker}, we display an environment resembling a solar 
  active 
  region comprising a multitude of magnetic bright points with flux tubes
  emerging from the photosphere. 
  Although the magnetic field spreads out to fill the corona, the merger is
  far from uniform and this presents an opportunity to explore the 
  dynamics between, and within, neighboring flux tubes.

  All of these examples are in magnetohydrostatic equilibrium for the 
  stratified solar atmosphere, with positive plasma pressure and density 
  throughout and low plasma-$\beta$ in the corona, while for $z\lesssim1\Mm$
  plasma-$\beta\gg1$.
  
  Typical footpoint field strength for each flux tube is near $100\mT$, but
  varies depending on the number and proximity of neighboring flux tubes. 
  Care in the field construction is required, with respect to the vertical and
  radial expansion factors, footpoint strength and axial proximity. 
  The magnetic field strength should not become so high as to 
  require plasma pressure or density to be negative in order to satisfy 
  the pressure balance.

  \section{Summary and Discussion} \label{sect:conc}
  We have solved analytically the time-independent MHD equation of 
  momentum for a configuration of
  multiple open magnetic flux tubes in magnetohydrostatic 
  equilibrium embedded within a solar stratified atmosphere, with realistic
  parameters for plasma pressure, 
  density, temperature, and magnetic field strength.
  The equilibrium is maintained through inclusion of appropriate 
  horizontal balancing forces, which we have also identified and calculated.
  It may be argued that inclusion even of an appropriate choice of
  balancing forces undermines the relationship between $\vect{B}, p$ and 
  $\rho$, permitting any arbitrary atmosphere to be constructed.
  This would negate comparison with the real solar atmosphere. 
  However, in our model the atmosphere is also constrained by observational
  comparison.
  We restrict the balancing forces to be the minimal requirement
  to solve the system, and so the atmospheric adjustments remain largely 
  determined by the applied magnetic field. 

  This is a significant advance in achieving realistic modeling of the 
  magnetized solar atmosphere.
  The solutions may also have application in other astrophysical
  environments{\apjrev{, such as sunspots or magnetized atmosphere in
  gas giants}}. 
  This advances the possibilities for analysis and numerical simulation of
  systems with background equilibria or quasi-equilibria.
  In particular, the existence of non-axisymmetric inhomogeneous 
  configurations facilitates research of non-linear interactions
  between neighboring flux tubes.
  The presence of converging field lines between flux tubes may lead to 
  interesting dynamics.
 
  The model extends from the solar photosphere to the solar corona,
  incorporating
  the temperature minimum and the TR. 
  There is some scale-independence to the model, in the sense that it could 
  apply to a single flux tube emerging from a magnetic 
  bright point or to more extended surface areas including flux tube 
  networks in an active solar region.
  In the smaller scales, such as magnetic bright points, the flux
  tubes environment is far from force free.
 
  There are two important analytical results.
  We have identified a sufficient condition to have a force-free steady
  solution for a
  magnetic field configuration subject to external vertical gravitation, as
  specified in Equation\,\eqref{eq:suff} Appendix\,\ref{sect:curl} 
  \begin{equation*}
    \frac{\partial B_x}{\partial y}
    =
    \frac{\partial B_y}{\partial x}
    \quad{\rm and}\quad
    \frac{\partial B_z}{\partial y}
    \frac{\partial B_x}{\partial z}
    =
    \frac{\partial B_z}{\partial x}
    \frac{\partial B_y}{\partial z}.
  \end{equation*}
  {{This result is consistent with that of \citet{BCLow85}, applying a
  general magnetic field of the form
  $$
    \vect{B} = \left(
      \frac{\partial \phi}{\partial x},
      \frac{\partial \phi}{\partial y},
      \psi
      \right),
  $$
  where $\phi$ and $\psi$ are scalar functions.
  The former condition in Equation\,\eqref{eq:suff} is immediately 
  satisfied, and the latter is equivalently specified by
  $$
    \frac{\partial^2 \phi}{\partial x\partial z}\frac{\partial \psi}{\partial y}
    =\frac{\partial^2 \phi}{\partial y\partial z}\frac{\partial \psi}{\partial x},
  $$ 
  for which the general solution is $\psi(x,y,z) = \Psi(z,\partial\phi/\partial z)$.
  This leads to the general solution of \citet{BCLow85} for the plasma pressure
  of the   form
  \begin{equation}
    \label{eq:low}
    p + \frac{1}{2\mu_0}\Psi^2 -  \frac{1}{\mu_0}\int^{\partial\phi/\partial z}
    \Psi(z,u)\dd u = p_0(z). 
  \end{equation}
  $p_0(z)$ is an arbitrary function ($p_b\vv$ in our model). 
  The term under the integral is typically non-linear without a ready analytic
  solution.
  \citet{BCLow85} found a particular solution with 
  $\Psi=\alpha \partial\phi/\partial z$. 
  Summing magnetic structures, as we have done in 
  Section\,\ref{sect:multi}, \citet{BCLow85} constructed a 
  non-axisymmetric multi-nodal configuration.

  In our model 
  \begin{equation}\label{eq:phi}
    \phi = \frac{\leftidx{^m}S\fO^2}{2\BO}\frac{\partial \BO}{\partial z}
           \leftidx{^m}\GO\,\,{\rm and}\,\,\psi 
         = \leftidx{^m}S\BO^2\leftidx{^m}\GO
         + \bc \neq \alpha \partial\phi/\partial z,
  \end{equation}
  so that, for the single flux tube, our model yields a new {{explicit}} solution
  to 
  Equation\,\eqref{eq:low}.
  The general solution of \citet{BCLow85} requires Equation\,\eqref{eq:suff} be 
  satisfied for Equation\,\eqref{eq:delP}}}, i.e., force-free.
  We have not verified whether this is a necessary condition, such that 
  failure to satisfy this condition would exclude the possibility of a 
  force-free steady solution.
  Here we introduce a new set of {{explicit}} solutions, 
  applicable when the latter condition in Equation\,\eqref{eq:suff} is 
  satisfied for the net magnetic forces applying to 
  Equation\,\eqref{eq:momm}. 
  These solutions allow general configurations for the magnetic field.
  Superpositions of various magnetic field configurations may be combined.
  Provided the former equality in Equation\eqref{eq:suff} is satisfied for each
  individual configuration, an analytic solution to Equation\,\eqref{eq:momm} 
  exists for $p_b$ and $\rho_b$, with $\vect{F}_{\rm bal}$ minimally defined by
  Equation\,\eqref{eq:Fbalsum}.

  Note that the former equality {{in Equation\eqref{eq:suff}}} 
  is equivalent to the condition $J_z = 0$,
  the vertical component of background current density
  and corresponds to an untwisted magnetic field configuration. 
  Our background field is not current-free.
  In the axisymmetric model, applying for the single flux tube, the background
  current is purely azimuthal. 
  For the multiple flux tubes, the background
  horizontal current includes radial and 
  azimuthal components.
  Let us stress that this does not exclude $J_z$ for the perturbed system.

  \citet{BCLow85} took the approach of identifying the conditions in
  Equation\,\eqref{eq:suff} and constructing a magnetic field to satisfy these. 
  His solution was applied to an exponential model background pressure, which
  approximates a coronal atmosphere.
  Our approach is to seek a construction for the magnetic field, 
  which is sufficiently flexible to adapt to the observed 
  magnetic structures and which matches 
  the more realistic solar atmosphere modeled by \citet{VAL81} and
  \citet{MTW75}, and would permit an analytic solution.
  We are motivated by physical considerations, such as controlling the
  radial expansion of the flux tubes distinctly for the chromosphere and for
  the corona.
  Therefore, the latter condition in Equation\,\eqref{eq:suff} has been used to
  identify those
  terms, which can only be accommodated by including balancing forces. 
  Although we approximate the magnetized background atmosphere as
  steady, both of these approaches have limitations.
  Particularly, in the chromosphere the atmosphere is neither static nor
  force free. 
  Many other physical forces may play a significant role, such as radiation, 
  partial ionization effects and thermal conduction.
  Nevertheless, this is a considerable advance toward modeling effectively
  some part of this complex and dynamic system.

  For our self-similar flux
  tube model, 
  we have verified that it is not possible to construct a field of
  neighboring flux tubes with a steady solution in the
  absence of balancing forces. 
  These forces can be identified and calculated, so that the 
  background be in
  quasi-equilibrium.
  It would be interesting to  investigate
  whether this approach could be extended to magnetic fields with twist, 
  such that the former condition in Equation\,\eqref{eq:suff} is not satisfied,
  but that balancing forces might yet be identified.
 
  An alternative approach could be to extend the analysis of 
  \citet{BCLow85} and seek a magnetic field that satisfies
  completely Equation\,\eqref{eq:suff} by construction. 
  From the solenoidal condition 
  \begin{equation}
   \frac{\partial^2 \phi}{\partial x^2}+
   \frac{\partial^2 \phi}{\partial y^2}+
   \frac{\partial}{\partial z}
   \left(
   f(z)\frac{\partial \phi}{\partial z}
   \right)
  \end{equation}
  we seek a solution with $f$ and $\phi$ in a form that may be adapted to match 
  the observed magnetic structures of the solar atmosphere.
  Finding such a solution is a considerable challenge and there is no guarantee
  that such configurations should realistically model open flux tubes or loops.
  While we will consider this approach in the future, it is well beyond the 
  current aims: the 
  construction of currently observable complex loop structures
  with the ultimate goal of investigating {\emph{wave propagation}} and
  {\emph{wave energy transport/coupling}} in such complex systems. 
  It is not our intention here to model loop or active region dynamics, 
  which involve evolution of the background due to processes, such as 
  reconnection, field relaxation, etc.

  In this article, we have restricted our examples to systems of open flux tubes
  of the same polarity.
  In Appendix\,\ref{sect:soln}, the derivation is also valid for
  solutions involving opposite polarity{{. Indeed, the constants
  $\leftidx{^m}S$ need not be identical, taking any set of real values subject
  to the constraint that plasma pressure and density remain physical}}.
  This could add further curvature for the magnetic field to the non-trivial
  field curvature that this article describes between flux tubes.
  A further, though very challenging, improvement would be to include 
  torsional components to the flux tubes, with $B_\phi\neq0$.

  Simulations using alternate steady background configurations with single and 
  multiple flux tubes, will help identify the extent to which the
  interactions between magnetic flux tubes amplify or dampen the transport of
  energy in the lower solar atmosphere. 
  Also analytical investigation of the various equilibrium conditions could 
  advance our understanding of the structure and forces acting in this solar
  region.
  
  \section*{Acknowledgements}
  F.A.G. is supported by STFC Grant R/131168-11-1. 
  R.E. is gratefull to the NSF, Hungary (OTKA, Ref. No. K83133) and 
  acknowledges M. K\'eray for patient encouragement. 
  The authors acknowledge the NumPy, SciPy \citep{jones2001}, 
  Matplotlib \citep{hunter2007} and MayaVi2 citep{ramachandran2011} Python 
  projects for providing the computational tools to analyze the data.
  We also thank Stuart Mumford (python), Stephen Chaffin and
  Alastair Williamson (useful discussions),
  and B.\,C.\,Low for 
  constructive comment on magnetohydrostatics. 
  {\apjrev{We thank the anonymous referee for helpful suggestions to improve 
  the clarity of the manuscript.}}

\appendix
\section{Solution to background static equilibrium }\label{sect:soln}
  \subsection{Non-vanishing Terms in the Curl of $\nabla p$}\label{sect:curl}
  If we consider Equation\,\eqref{eq:curlP} 
  for a general magnetic field $\vect{B}$ subject to 
  gravity acting only along the vertical direction, we require 
  \begin{equation}\label{eq:curlPa}
    \nabla\times\nabla P= \nabla\times 
    \left(
      \rho g\vect{\hat{z}} 
      +
      \frac{
      \left(
        \vect{B}\cdot\nabla
      \right)
      \vect{B}
      }{\mu_0} 
      -\nabla \frac{|\vect{B}|^2}{2\mu_0}
    \right) = \vect{0}.
  \end{equation}
  On the left-hand side the pressure is a scalar, so this term vanishes.
  The magnetic pressure term as a scalar also has vanishing curl, so for the 
  magnetic tension and gravitation we require
  \begin{align}
  \label{eq:curlx}
    \frac{\partial}{\partial y}
    \left[
      (\vect{B}\cdot\nabla)\frac{B_z}{\mu_0} + \rho g
    \right]
    -
    \frac{\partial}{\partial z}
    \left[
      (\vect{B}\cdot\nabla)\frac{B_y}{\mu_0}
    \right] 
    &=&0,
    \\
  \label{eq:curly}
    \frac{\partial}{\partial z}
    \left[
      (\vect{B}\cdot\nabla)\frac{B_x}{\mu_0}
    \right] 
    -
    \frac{\partial}{\partial x}
    \left[
      (\vect{B}\cdot\nabla)\frac{B_z}{\mu_0} + \rho g
    \right]
    &=&0,
    \\
  \label{eq:curlz}
    \frac{\partial}{\partial x}
    \left[
      (\vect{B}\cdot\nabla)\frac{B_y}{\mu_0}
    \right] 
    -
    \frac{\partial}{\partial y}
    \left[
      (\vect{B}\cdot\nabla)\frac{B_x}{\mu_0}
    \right]
    &=&0.
  \end{align}
  From 
  Equation\,\eqref{eq:curlz} we obtain
  \begin{align}\label{eq:curlxy}
    \frac{\partial}{\partial x}
    \left[
      (\vect{B}\cdot\nabla)B_y)
    \right] 
    -
    \frac{\partial}{\partial y}
    \left[
      (\vect{B}\cdot\nabla)B_x)
    \right] = 0
    \Rightarrow
    \left[
    \vect{B}\cdot\nabla
    -
    \frac{\partial B_z}{\partial z}
    \right]
    \left(
    \frac{\partial B_y}{\partial x}
    -
    \frac{\partial B_x}{\partial y}
    \right)
    +
    \frac{\partial B_z}{\partial x}
    \frac{\partial B_y}{\partial z}
    -
    \frac{\partial B_z}{\partial y}
    \frac{\partial B_x}{\partial z}
    =0
    . 
  \end{align}
  A solution to Equation\,\eqref{eq:curlxy} exists if 
  \begin{equation}\label{eq:suff}
    \frac{\partial B_x}{\partial y}
    =
    \frac{\partial B_y}{\partial x}
    \quad{\rm and}\quad
    \frac{\partial B_z}{\partial y}
    \frac{\partial B_x}{\partial z}
    =
    \frac{\partial B_z}{\partial x}
    \frac{\partial B_y}{\partial z}
  \end{equation}
  Equation\,\eqref{eq:suff} is a sufficient condition to satisfy 
  Equation\,\eqref{eq:curlz}.
  Differentiating Equation\,\eqref{eq:curlx} with respect to $x$ and 
  Equation\,\eqref{eq:curly} with respect to $y$ and then
  summing them, we are left only
  with the $z$-derivative of Equation\,\eqref{eq:curlxy}.
  Again we obtain the relations in Equation\,\eqref{eq:suff} as a sufficient
  condition to fully satisfy Equation\,\eqref{eq:curlPa}.
  Indeed, this may be a necessary condition
  for a steady magnetic field 
  within a vertical gravity field \citep{Gabriel76,GL98}, 
  but we have not verified this. 

  In the case of the self-similar construction for a single flux tube defined  
  by Equation\,\eqref{eq:Bxyz}, {\apjrev{both conditions in Equation\,\eqref{eq:suff} are
  }}satisfied.
  However, for a pair of flux tubes denoted by $\leftidx{^m}{\vect{B}_b}$
  and $\leftidx{^n}{\vect{B}_b}$,
  where $x_i\neq x_j$ or $y_i\neq y_j$,
  the latter condition in Equation\,\eqref{eq:suff} is not satisfied
  for the cross terms
  \begin{equation}\label{eq:fail}
    \frac{\partial \leftidx{^m}{B_{bz}}}{\partial y}
    \frac{\partial \leftidx{^n}{B_{bx}}}{\partial z}
    \neq                                 
    \frac{\partial \leftidx{^m}{B_{bz}}}{\partial x}
    \frac{\partial \leftidx{^n}{B_{by}}}{\partial z}.
  \end{equation}
  In general configurations of vertical flux tube pairs may 
  include derivative terms for which $y_ix_j\neq x_iy_j$, failing to 
  satisfy Equation\,\eqref{eq:suff}.
  We have not identified an alternative construction in 
  which this pairwise interaction can satisfy Equation\,\eqref{eq:suff}.
  It may be that for an asymmetric field magnetohydrostatic equilibrium cannot
  exist in the absence of balancing forces, but we have not verified this.

  From Equation\,\eqref{eq:fail} the non-vanishing terms in Equation\,\eqref{eq:curlPa}
  can be identified.
  A sufficient requirement to equate the right-hand side
  to $\vect{0}$ will thus be to include a balancing force inside the brackets 
  {\small{\begin{equation}\label{eq:Fbal}
    -\frac{\leftidx{^m}{B_{bz}}}{\mu_0}\frac{\partial \leftidx{^n}{B_{bx}}}{\partial z}\vect{\hat{x}}
    -\frac{\leftidx{^n}{B_{bz}}}{\mu_0}\frac{\partial \leftidx{^m}{B_{bx}}}{\partial z}\vect{\hat{x}}
    -\frac{\leftidx{^m}{B_{bz}}}{\mu_0}\frac{\partial \leftidx{^n}{B_{by}}}{\partial z}\vect{\hat{y}}
    -\frac{\leftidx{^n}{B_{bz}}}{\mu_0}\frac{\partial \leftidx{^m}{B_{by}}}{\partial z}\vect{\hat{y}}.
  \end{equation}}}It follows that it is sufficient to modify  
  the interaction terms in 
  Equation\,\eqref{eq:delPab} to
  \begin{eqnarray}\label{eq:delPa} 
    \nabla \leftidx{^{mn}}{p_b\hh}  
  + \nabla\left(\leftidx{^m}{\vect{B}_b}\cdot
    \frac{\leftidx{^n}{\vect{B}_b}}{\mu_0}\right) 
  - \left(\frac{\leftidx{^m}{\vect{B}_b}}{\mu_0}\cdot
    \nabla\right)\leftidx{^n}{\vect{B}_b} 
  - \left(\frac{\leftidx{^n}{\vect{B}_b}}{\mu_0}\cdot
    \nabla\right)\leftidx{^m}{\vect{B}_b}\,
    &&
    \\
    \nonumber
    +\frac{\leftidx{^m}{B_{bz}}}{\mu_0}\frac{\partial \leftidx{^n}{B_{bx}}}{\partial z}\vect{\hat{x}}
    +\frac{\leftidx{^n}{B_{bz}}}{\mu_0}\frac{\partial \leftidx{^m}{B_{bx}}}{\partial z}\vect{\hat{x}}\,
    +\frac{\leftidx{^m}{B_{bz}}}{\mu_0}\frac{\partial \leftidx{^n}{B_{by}}}{\partial z}\vect{\hat{y}}
    +\frac{\leftidx{^n}{B_{bz}}}{\mu_0}\frac{\partial \leftidx{^m}{B_{by}}}{\partial z}\vect{\hat{y}}
    - \leftidx{^{mn}}{\rho_b\hh}g\vect{\hat{z}}
    &=&\vect{0}.
  \end{eqnarray} 
  In this form we can now follow 
  Appendix\,\ref{sect:pbal} to solve for $p$ and $\rho$.
  With plasma pressure and density thus modified, the equality in
  Equation\,\eqref{eq:delP} for the pressure balance is no longer valid.
  This will be restored by adding to the right-hand side the sum of
  forces $\leftidx{^{mn}}{\vect{F}}$ matching the net force applying on
  $\leftidx{^n}{\vect{B}_b}$ due to 
  $\leftidx{^m}{\vect{B}_b}$. Explicitly
  \begin{eqnarray}\label{eq:Fij}
   \leftidx{^{mn}}{\vect{F}}\, &=& -\frac{\leftidx{^m}{B_{bz}}}{\mu_0}
   \frac{\partial \leftidx{^n}{B_{bx}}}{\partial z}\vect{\hat{x}}
                   -  \frac{ \leftidx{^m}{B_{bz}}}{\mu_0}
   \frac{\partial \leftidx{^n}{B_{by}}}{\partial z}\vect{\hat{y}}
    \\
    \nonumber
    &=&
    \left[\frac{(x-\leftidx{^n}x )\vect{\hat{x}}
      +(y-\leftidx{^n}y )\vect{\hat{y}}}{\mu_0}
    \right]
    \leftidx{^n}S
    \leftidx{^n}\GO 
    \left[
    \leftidx{^m}S
    \BO^2\leftidx{^m}\GO 
    +
    \bc
    \right]
    \left[
      \left(
      1-\frac{2\leftidx{^n}f ^2}{\fO^2}
      \right)
      \left(
        \frac{\partial \BO}{\partial z}
      \right)^2
      +\BO\frac{\partial^2 \BO}{\partial z^2}
    \right]
  \end{eqnarray}
   
  \subsection{Plasma Pressure and Density Adjustment}\label{sect:pbal}
  \subsubsection{Basic Quantities and Derivatives}\label{sect:basic}
  Listed here are
  the various
  derivatives from the expressions for a single flux tube, which will be 
  required in the calculations.
  \begin{equation*}
    \frac{\partial \leftidx{^m}f }{\partial x}=
    \frac{(x-\leftidx{^m}x )\BO}{
    \leftidx{^m}r \, }
    ,\,\quad 
    \frac{\partial \leftidx{^m}f }{\partial y}
    =
    \frac{(y-\leftidx{^m}y )\BO}{
    \leftidx{^m}r \, }
    ,\,\quad 
    \frac{\partial \leftidx{^m}f }{\partial z}=
    {\leftidx{^m}r \, }{}
    \frac{\partial \BO}{\partial z},\quad
    \frac{\partial \leftidx{^m}\GO }{\partial \leftidx{^m}f \,}=-\frac{2\leftidx{^m}f \leftidx{^m}\GO }{\fO^2}, 
  \end{equation*}
  \begin{equation*}
    \frac{\partial \leftidx{^m}\GO }{\partial x}= 
    -\frac{2(x-\leftidx{^m}x )\BO^2\leftidx{^m}\GO }{\fO^2 },\quad 
    \frac{\partial \leftidx{^m}\GO }{\partial y}=
    -\frac{2(y-\leftidx{^m}y )\BO^2\leftidx{^m}\GO }{\fO^2 },\quad 
    \frac{\partial \leftidx{^m}\GO }{\partial z}=
    -\frac{2 \leftidx{^m}f \leftidx{^m}\GO  \leftidx{^m}r }{\fO^2}
    \frac{\partial \BO}{\partial z}.
  \end{equation*}
  
  We will require the derivatives in these expressions as follows: 
  \begin{eqnarray}\label{eq:dbxdx}
    \frac{\partial \leftidx{^m}{B_{bx}}}{\partial x}
    &=& 
    \leftidx{^m}S{{\leftidx{^m}\GO \, }
    \left(
      \frac{2(x-\leftidx{^m}x )^2\BO^3}{\fO^2 }-\BO
    \right)
    \frac{\partial \BO}{\partial z}}
  \end{eqnarray}
  \begin{equation}
    \label{eq:dbydx}
    \frac{\partial \leftidx{^m}{B_{bx}}}{\partial y}
    = 
    \frac{\partial \leftidx{^m}{B_{by}}}{\partial x}
    = 
    \frac{2\leftidx{^m}S(x-\leftidx{^m}x )(y-\leftidx{^m}y )\BO^3\leftidx{^m}\GO }{\fO^2 }
    \frac{\partial \BO}{\partial z}
  \end{equation}
  \begin{eqnarray}\label{eq:dbxdz}
    \frac{\partial \leftidx{^m}{B_{bx}}}{\partial z}
    = 
    \leftidx{^m}S(x-\leftidx{^m}x )
    \leftidx{^m}\GO 
    \left(
      \frac{2 \leftidx{^m}f ^2 }{\fO^2}
      -1
    \right)
    \frac{\partial \BO}{\partial z}^2
    -\leftidx{^m}S
    (x-\leftidx{^m}x )
    \BO
    \leftidx{^m}\GO 
    \frac{\partial^2 \BO}{\partial z^2}
  \end{eqnarray}
  \begin{eqnarray}\label{eq:dbydy}
    \frac{\partial \leftidx{^m}{B_{by}}}{\partial y}
    = 
    \leftidx{^m}S{{\leftidx{^m}\GO \, }
    \left(
      \frac{2(y-\leftidx{^m}y )^2\BO^3}{\fO^2 }-\BO
    \right)
    \frac{\partial \BO}{\partial z}}
  \end{eqnarray}
  \begin{eqnarray}\label{eq:dbydz}
    \frac{\partial \leftidx{^m}{B_{by}}}{\partial z}
    = 
    \leftidx{^m}S(y-\leftidx{^m}y )
    \leftidx{^m}\GO 
    \left(
      \frac{2 \leftidx{^m}f ^2 }{\fO^2}
      -
      1
    \right)
    \frac{\partial \BO}{\partial z}^2
    -
    \leftidx{^m}S(y-\leftidx{^m}y )
    \BO
    \leftidx{^m}\GO 
    \frac{\partial^2 \BO}{\partial z^2}
  \end{eqnarray}  
  \begin{eqnarray}\label{eq:dbzdx}
    \frac{\partial \leftidx{^m}{B_{bz}}}{\partial x}
    &=& 
    -\leftidx{^m}S\frac{2(x-\leftidx{^m}x )\BO^4 \leftidx{^m}\GO }{\fO^2 },
  \end{eqnarray}
  \begin{eqnarray}\label{eq:dbzdy}
    \frac{\partial \leftidx{^m}{B_{bz}}}{\partial y}
    &=&
    -\leftidx{^m}S\frac{2(y-\leftidx{^m}y )\BO^4 \leftidx{^m}\GO }{\fO^2 },
  \end{eqnarray}
  \begin{eqnarray}\label{eq:dbzdz}
    \frac{\partial \leftidx{^m}{B_{bz}}}{\partial z}
    &=&
    {2\leftidx{^m}S\BO\leftidx{^m}\GO \, }
    \left(
      1-\frac{\leftidx{^m}f ^2}{\fO^2}
    \right)
    \frac{\partial \BO}{\partial z}
  \end{eqnarray}
  
  \subsubsection{Balancing Plasma Pressure and Density for Single Flux Tube}

  Each single flux tube in isolation as prescribed by Equation\,\eqref{eq:Bxyz} 
  has axial symmetry.
  For convenience we retrace the solution in cylindrical coordinates as 
  described in \GFME, but here applying a constant vertical ambient
  background field.
  We also include the constant $\leftidx{^m}S=\pm1$, allowing alternative polarity
  to be considered for each flux tube. 
  \begin{eqnarray*}
    \frac{\partial \leftidx{^m}{p_b\hh}}{\partial r}&=&
    -\frac{\partial }{\partial r} 
    \frac{|\leftidx{^m}{\vect{B}_b}|^2}{2\mu_0}
    +\frac{\leftidx{^m}{B_{br}}}{\mu_0}\frac{\partial \leftidx{^m}{B_{br}}}{\partial r} 
    +\frac{\leftidx{^m}{B_{bz}}}{\mu_0}\frac{\partial \leftidx{^m}{B_{br}}}{\partial z} 
    \\
    \nonumber
    &=&
    -\frac{\partial }{\partial r} 
    \frac{|\leftidx{^m}{\vect{B}_b}|^2}{2\mu_0}
    +\frac{\partial }{\partial r} 
    \frac{\leftidx{^m}{B_{br}}^2}{2\mu_0}
    -
    \frac{(\leftidx{^m}S\BO^2\leftidx{^m}\GO +\bc)}{\mu_0}
    \frac{\partial }{\partial z}
    \left(
    \leftidx{^m}S\leftidx{^m}r \BO\leftidx{^m}\GO \frac{\partial \BO }{\partial z}
    \right)
    \\
    \nonumber
    &=&
    -\frac{\partial }{\partial r} 
    \frac{\leftidx{^m}{B_{bz}}^2}{2\mu_0}
    -
   \frac{ \leftidx{^m}S^2\BO \leftidx{^m}f \leftidx{^m}\GO ^2}{\mu_0}
    \left(
    \frac{\partial \BO }{\partial z}
    \right)^2
    +
    \frac{2\leftidx{^m}S^2\BO \leftidx{^m}f ^3\leftidx{^m}\GO ^2}{\mu_0\fO^2}
    \left(
    \frac{\partial \BO }{\partial z}
    \right)^2
    -
    \frac{\leftidx{^m}S\bc \leftidx{^m}f \leftidx{^m}\GO}{\mu_0}   
    \frac{\partial^2 \BO }{\partial z^2}
    \\
    \nonumber
    &+&\left[
    \frac{2\leftidx{^m}S\bc \leftidx{^m}f ^3\leftidx{^m}\GO }{\mu_0\BO\fO^2}
    -\frac{\leftidx{^m}S\bc \leftidx{^m}f \leftidx{^m}\GO }{\mu_0\BO}
    \right]
    \left(
    \frac{\partial \BO }{\partial z}
    \right)^2
    -
    \frac{\leftidx{^m}S^2\BO^2\leftidx{^m}f \leftidx{^m}\GO ^2}{\mu_0}
    \frac{\partial^2 \BO }{\partial z^2}
    \\
    &=&
    -\frac{\partial }{\partial r} 
    \frac{\leftidx{^m}{B_{bz}}^2}{2\mu_0
    }-
    \frac{\leftidx{^m}S^2}{2\mu_0}
    \frac{\partial }{\partial r}
    \left(
    \leftidx{^m}f \leftidx{^m}\GO  
    \frac{\partial \BO}{\partial z}
    \right)^2
    +
    \frac{\partial }{\partial r}
    \left(
    \frac{\leftidx{^m}S^2\BO\fO^2\leftidx{^m}\GO ^2 }{4\mu_0}
    \frac{\partial^2 \BO}{\partial z^2}
    \right)
    \\
    \nonumber
    &+&
    \frac{\partial }{\partial r}
    \left(
    \frac{\leftidx{^m}S\bc\fO^2\leftidx{^m}\GO  }{2\mu_0\BO}
    \frac{\partial^2 \BO}{\partial z^2}
    \right)
    -
    \frac{\partial }{\partial r}
    \left(
    \frac{\leftidx{^m}S\bc \leftidx{^m}f ^2\leftidx{^m}\GO }
    {\mu_0\BO^2}
    \left[
    \frac{\partial \BO}{\partial z}
    \right]^2
    \right)
    -
    \frac{\partial }{\partial r}
    \left(
    \frac{\leftidx{^m}S\bc\fO^2\leftidx{^m}\GO }
    {2\mu_0\BO^2}
    \left[
    \frac{\partial \BO}{\partial z}
    \right]^2
    \right).
  \end{eqnarray*}
  Integrating with respect to $r$ we obtain a solution for $\leftidx{^m}p_b\hh$
  as
  \begin{equation}\label{eq:phfull}
    \leftidx{^m}{p_b\hh}\,
    =
    -\frac{|\leftidx{^m}{\vect{B_b}}|^2}{2\mu_0}
    -
    \frac{\leftidx{^m}S\bc\leftidx{^m}\GO }
    {\mu_0\BO^2}
    \left[
      \leftidx{^m}f ^2+\frac{\fO^2}{2}
    \right]
    \left(
      \frac{\partial \BO}{\partial z}
    \right)^2
    +
    \frac{\leftidx{^m}S^2\BO\fO^2\leftidx{^m}\GO ^2 }{4\mu_0}
    \frac{\partial^2 \BO}{\partial z^2}
    +
    \frac{\leftidx{^m}S\bc\fO^2\leftidx{^m}\GO  }{2\mu_0\BO}
    \frac{\partial^2 \BO}{\partial z^2}.
  \end{equation}
  This must be matched by the solution obtained by solving the $z$-component of
  the pressure balance equation so 
  \begin{eqnarray*}
    \leftidx{^m}{p_b\hh}\,&=&
    \int
      \leftidx{^m}{\rho_b\hh} g\dd z 
      - \frac{\leftidx{^m}{B_{br}}^2}{2\mu_0}
      +\int \frac{\leftidx{^m}{B_{br}}}{\mu_0}
       \frac{\partial \leftidx{^m}{B_{bz}}}{\partial r}
    \dd z 
    =
    - \frac{\leftidx{^m}{B_{bz}}^2}{2\mu_0}
    +\int
      \frac{\leftidx{^m}{B_{bz}}}{\mu_0}\frac{\partial \leftidx{^m}{B_{br}}}{\partial z}
    \dd r 
    \\
    \nonumber
    \Rightarrow\leftidx{^m}{\rho_b\hh} g
    &=&
    \frac{\partial }{\partial z} 
    \frac{\leftidx{^m}{B_{br}}^2}{2\mu_0}
    -\frac{\partial }{\partial z}
    \frac{\leftidx{^m}{B_{bz}}^2}{2\mu_0} 
    - \frac{\leftidx{^m}{B_{br}}}{\mu_0}\frac{\partial \leftidx{^m}{B_{bz}}}{\partial r} 
    +\frac{\partial }{\partial z} 
    \int 
      \frac{\leftidx{^m}{B_{bz}}}{\mu_0}\frac{\partial \leftidx{^m}{B_{br}}}{\partial z}
    \dd r 
  \end{eqnarray*}
  \begin{eqnarray*}
    \leftidx{^m}{\rho_b\hh} g
    &=&
    \frac{\partial }{\partial z}
    \left(
      \cancel{
      \frac{\leftidx{^m}S^2\leftidx{^m}f ^2\leftidx{^m}\GO ^2 }{2\mu_0}
      \left[
        \frac{\partial \BO}{\partial z}
      \right]^2
      }
      -
      \frac{\leftidx{^m}S^2\BO^4\leftidx{^m}\GO ^2}{2\mu_0}
    \right)
    -
    \frac{\partial }{\partial z}
    \left(
      \frac{\leftidx{^m}S\bc\BO^2\leftidx{^m}\GO}{\mu_0} 
    \right)
    \\
    \nonumber
    &-& 
    \frac{\partial }{\partial z}
    \left(
      \cancel{
      \frac{\leftidx{^m}S^2\leftidx{^m}f ^2\leftidx{^m}\GO ^2 }{2\mu_0}
      \left[
        \frac{\partial \BO}{\partial z}
      \right]^2
      }  
    \right) 
    +
    \frac{\partial }{\partial z}
    \left(
      \frac{\leftidx{^m}S^2\BO\fO^2\leftidx{^m}\GO ^2 }{4\mu_0}
      \frac{\partial^2 \BO}{\partial z^2}
      +
      \frac{\leftidx{^m}S\bc\fO^2\leftidx{^m}\GO  }{2\mu_0\BO}
      \frac{\partial^2 \BO}{\partial z^2}
    \right)
    \\
    \nonumber
    &-&
    \frac{\partial }{\partial z}
    \left(
      \frac{\leftidx{^m}S\bc \leftidx{^m}\GO }
      {\mu_0\BO^2}
      \left(
        \leftidx{^m}f ^2+
        \frac{\fO^2}
        {2}
      \right)
      \left[
        \frac{\partial \BO}{\partial z}
      \right]^2
    \right)
    + 
    \frac{\leftidx{^m}S^2\leftidx{^m}f \leftidx{^m}\GO}{\mu_0} 
    \frac{\partial \BO}{\partial z}
    \frac{\partial }{\partial r}
    \left(
      \BO^2\leftidx{^m}\GO 
    \right) 
  \end{eqnarray*}
  
  \begin{eqnarray}\label{eq:rho}
    \leftidx{^m}{\rho_b\hh}\, &=&
    \nonumber
     \frac{\leftidx{^m}S^2
     \leftidx{^m}\GO ^2}{\mu_0g}\left(\frac{\fO^2}{4}-\leftidx{^m}f ^2\right) 
    \frac{\partial \BO}{\partial z}
    \frac{\partial^2 \BO}{\partial z^2}
    -
      \frac{3\leftidx{^m}S\bc\leftidx{^m}\GO }{\mu_0g\BO^2}
      \left(
        \frac{\fO^2}{2}+\leftidx{^m}f ^2
      \right) 
    \frac{\partial \BO}{\partial z}
    \frac{\partial^2 \BO}{\partial z^2}
    \\
    \nonumber
    &+&
    \left[
      \frac{\leftidx{^m}S\bc \leftidx{^m}\GO }
      {\mu_0g\BO^3}
      \left(
        \frac{2 \leftidx{^m}f ^4}
        {\fO^2}+\fO^2+\leftidx{^m}f ^2
      \right)
    \right]
    \left[
      \frac{\partial \BO}{\partial z}
    \right]^3
    +
      \frac{2\leftidx{^m}S\bc\BO\leftidx{^m}\GO}{\mu_0g} \left(\frac{\leftidx{^m}f ^2}{\fO^2}-1\right)
    \frac{\partial \BO}{\partial z}
    \\
    &+&
    \left[
      \frac{\leftidx{^m}S^2\BO\fO^2\leftidx{^m}\GO ^2 }{4\mu_0g}
      +
      \frac{\leftidx{^m}S\bc\fO^2\leftidx{^m}\GO  }{2\mu_0g\BO}
    \right]
    \frac{\partial^3 \BO}{\partial z^3}
     -
     \frac{2\leftidx{^m}S^2\BO^3\leftidx{^m}\GO ^2
   }{\mu_0g} \frac{\partial \BO}{\partial z}
    .
  \end{eqnarray}
  The distribution for the plasma density can therefore be obtained by 
  dividing Equation\,\eqref{eq:rho} by $g$. 
  For our example we apply a constant $g$ for simplicity due to the small 
  variation over the vertical domain of our model, but the solution is equally
  valid with variable gravity  $g=g(z)$.
  However, it is not suitable for including self-gravity due to the
  horizontal 
  fluctuations in the plasma density. 
  This would arguably be very small in comparison to the global solar gravity
  for the scales we are considering.  
  {\subsubsection{{Plasma Pressure and Density from Pairwise Interactions}}}
%
  We require $\leftidx{^{mn}}{p_b\hh}$, the pressure deviation in response to
  the force 
  exerted by flux tube configuration $\leftidx{^m}{\vect{B}_b}$ 
  on $\leftidx{^n}{\vect{B}_b}$ and 
  vice versa.
  {{Taking advantage of the equality in Equation\,\eqref{eq:dbydx}
  the $y$-derivatives in Equation\,\eqref{eq:dpdx} can be transposed and the 
  tension terms cancel directly to yield
  }}
  {{
  \begin{eqnarray*}
    \frac{\partial \leftidx{^{mn}}{p_b\hh}}{\partial x}
    &=& 
    -\frac{\partial }{\partial x}
    \left(
    \frac{\leftidx{^m}{\vect{B}_b}\cdot\leftidx{^n}{\vect{B}_b}}{\mu_0}
    \right)
     +
      \frac{\leftidx{^m}{B_{bx}}}{\mu_0}\frac{\partial \leftidx{^n}{B_{bx}}}{\partial x}
      +\frac{\leftidx{^n}{B_{bx}}}{\mu_0}\frac{\partial \leftidx{^m}{B_{bx}}}{\partial x}
     +
      \frac{\leftidx{^m}{B_{by}}}{\mu_0}\frac{\partial \leftidx{^n}{B_{by}}}{\partial x}
      +\frac{\leftidx{^n}{B_{by}}}{\mu_0}\frac{\partial \leftidx{^m}{B_{by}}}{\partial x}
    \\
    \nonumber
    &=& 
    -\frac{\partial }{\partial x}
    \left(
    \frac{\leftidx{^m}{{B}_{bz}}\leftidx{^n}{{B}_{bz}}}{\mu_0}
    \right).
  \end{eqnarray*} 
  Integrating with respect to $x$ we thus 
  obtain}}
  {{
  \begin{eqnarray}\label{eq:apx}
    \leftidx{^{mn}}{p_b\hh}\,&=&
    -\frac{\leftidx{^m}{{B}_{bz}}\leftidx{^n}{{B}_{bz}}}{\mu_0}
  \end{eqnarray}
  }}plus an arbitrary function constant in $x$.
  Solving Equation\,\eqref{eq:dpdy} for the $y$-component 
  recovers the identical solution, so 
  any additional terms can have only $z$-dependence and are fully accounted for
  within the hydrostatic background terms $p_b\vv$ and $\rho_b\vv$.
  The solution to Equation\,\eqref{eq:dpdz} must also match so 
  \begin{eqnarray*}
    \leftidx{^{mn}}{p_b\hh}\, 
    &=& 
    \int
      \leftidx{^{mn}}{\rho_b\hh}g\nonumber 
      +(\leftidx{^m}{\vect{B}_b}\cdot\nabla) \frac{\leftidx{^n}{B_{bz}}}{\mu_0}
      + (\leftidx{^n}{\vect{B}_b}\cdot\nabla) \frac{\leftidx{^m}{B_{bz}}}{\mu_0}
    \dd z
    - \frac{\leftidx{^m}{\vect{B}_b}\cdot\leftidx{^n}{\vect{B}_b}}{\mu_0}
    =
    {{-\frac{\leftidx{^m}{{B}_{bz}}\leftidx{^n}{{B}_{bz}}}{\mu_0}}}
  \end{eqnarray*}
  \begin{eqnarray*}
    \Rightarrow \leftidx{^{mn}}{\rho_b\hh}g 
    &=&
    \nonumber
    \frac{\partial }{\partial z}\left(
    \frac{\leftidx{^m}{B_{bx}}\leftidx{^n}{B_{bx}}}{\mu_0}
    \right)
    +{{\frac{\partial }{\partial z}\left(
    \frac{\leftidx{^m}{B_{by}}\leftidx{^n}{B_{by}}}{\mu_0}
    \right)}}
      -(\leftidx{^m}{\vect{B}_b}\cdot\nabla) \frac{\leftidx{^n}{B_{bz}}}{\mu_0}
      - (\leftidx{^n}{\vect{B}_b}\cdot\nabla) \frac{\leftidx{^m}{B_{bz}}}{\mu_0}
  \end{eqnarray*}
{\small{
  \begin{eqnarray*}
    \mu_0 
    \leftidx{^{mn}}{\rho_b\hh}g 
    &=&
    \nonumber
    \frac{\partial }{\partial z}
    \left(
      \leftidx{^m}S\leftidx{^n}S(x-\leftidx{^m}x )(x-\leftidx{^n}x )
      \leftidx{^m}\GO 
      \leftidx{^n}\GO 
      \left[
        \BO
        \frac{\partial \BO}{\partial z}
      \right]^2
    \right)
    -
    \frac{\partial }{\partial z}
    \left(
      \leftidx{^m}{B_{bz}}\leftidx{^n}{B_{bz}}
      \right)
    -\leftidx{^m}{B_{bx}}\frac{\partial \leftidx{^n}{B_{bz}}}{\partial x}
    \\
    \nonumber
    &+&
    \frac{\partial }{\partial z}
    \left(
      \leftidx{^m}S\leftidx{^n}S(y-\leftidx{^m}y )(y-\leftidx{^n}y )
      \leftidx{^m}\GO 
      \leftidx{^n}\GO 
      \left[
        \BO
        \frac{\partial \BO}{\partial z}
      \right]^2
    \right)
    -\leftidx{^m}{B_{by}}\frac{\partial \leftidx{^n}{B_{bz}}}{\partial y}
    -
    \leftidx{^n}{B_{bx}}\frac{\partial \leftidx{^m}{B_{bz}}}{\partial x}
    -\leftidx{^n}{B_{by}}\frac{\partial \leftidx{^m}{B_{bz}}}{\partial y}
  \end{eqnarray*}
  \begin{eqnarray*}
    \ldots 
    &=&
    \nonumber
    2\leftidx{^m}S\leftidx{^n}S(x-\leftidx{^m}x )(x-\leftidx{^n}x )\BO
    \leftidx{^m}\GO 
    \leftidx{^n}\GO 
    \left[
      \frac{\partial \BO}{\partial z}
    \right]^3
    +
    2\leftidx{^m}S\leftidx{^n}S(x-\leftidx{^m}x )(x-\leftidx{^n}x )\BO^2
    \leftidx{^m}\GO 
    \leftidx{^n}\GO 
    \frac{\partial \BO}{\partial z}
    \frac{\partial^2 \BO}{\partial z^2}
    \\
    \nonumber
    &-&
    2\leftidx{^m}S\leftidx{^n}S(x-\leftidx{^m}x )(x-\leftidx{^n}x )\BO
    \frac{\leftidx{^m}f ^2}{\fO^2}
    \leftidx{^m}\GO 
    \leftidx{^n}\GO 
    \left[
      \frac{\partial \BO}{\partial z}
    \right]^3
    -
    2\leftidx{^m}S\leftidx{^n}S(x-\leftidx{^m}x )(x-\leftidx{^n}x )\BO
    \frac{\leftidx{^n}f ^2}{\fO^2}
    \leftidx{^m}\GO 
    \leftidx{^n}\GO 
    \left[
      \frac{\partial \BO}{\partial z}
    \right]^3
    \\
    \nonumber
    &+&
    2\leftidx{^m}S\leftidx{^n}S(y-\leftidx{^m}y )(y-\leftidx{^n}y )\BO
    \leftidx{^m}\GO 
    \leftidx{^n}\GO 
    \left[
      \frac{\partial \BO}{\partial z}
    \right]^3
    +
    2\leftidx{^m}S\leftidx{^n}S(y-\leftidx{^m}y )(y-\leftidx{^n}y )\BO^2
    \leftidx{^m}\GO 
    \leftidx{^n}\GO 
    \frac{\partial \BO}{\partial z}
    \frac{\partial^2 \BO}{\partial z^2}
  \end{eqnarray*}
  }}
  \begin{eqnarray}\label{eq:rhoijg}
    \leftidx{^{mn}}{\rho_b\hh}  
    \nonumber
    &=&
    \frac{2}{\mu_0g}\leftidx{^m}S\leftidx{^n}S
    (x-\leftidx{^m}x )(x-\leftidx{^n}x )
    \BO
    \leftidx{^m}\GO 
    \leftidx{^n}\GO 
    \left[
      \frac{\fO^2-\leftidx{^m}f ^2-\leftidx{^n}f ^2}{\fO^2}
      \left(
        \frac{\partial \BO}{\partial z}
      \right)^3
      +\BO
      \frac{\partial \BO}{\partial z}
      \frac{\partial^2 \BO}{\partial z^2}
    \right]
    \\
    \nonumber
    &+&
    \frac{2}{\mu_0g}\leftidx{^m}S\leftidx{^n}S(y-\leftidx{^m}y )(y-\leftidx{^n}y )\BO
    \leftidx{^m}\GO 
    \leftidx{^n}\GO 
    \left[
      \frac{\fO^2-\leftidx{^m}f ^2-\leftidx{^n}f ^2}{\fO^2}
      \left(
        \frac{\partial \BO}{\partial z}
      \right)^3
      +\BO
      \frac{\partial \BO}{\partial z}
      \frac{\partial^2 \BO}{\partial z^2}
    \right]
    \\
    \nonumber
    &+& 
      \frac{2}{\mu_0g} \leftidx{^m}S\leftidx{^n}S\BO^3\leftidx{^m}\GO \leftidx{^n}\GO 
    \frac{\partial \BO}{\partial z}
    \left[
      \frac{\BO^2(\leftidx{^n}x-\leftidx{^m}x )^2}{\fO^2}
      +\frac{\BO^2(\leftidx{^n}y-\leftidx{^n}y )^2}{\fO^2}-2
    \right]
    \\
    &+& 
    \frac{2}{\mu_0g}\bc\BO
    \frac{\partial \BO}{\partial z}
    \left[
      \frac{\leftidx{^m}S\leftidx{^m}\GO \leftidx{^m}f ^2}{\fO^2}
      -  \leftidx{^m}S\leftidx{^m}\GO 
      +
      \frac{\leftidx{^n}S\leftidx{^n}\GO \leftidx{^n}f ^2}{\fO^2}
      -  \leftidx{^n}S\leftidx{^n}\GO 
    \right]
    .
  \end{eqnarray}
  \bibliographystyle{apj}      
  \bibliography{fred_mnras}
  \label{lastpage}
%


\end{document}